\RequirePackage{lineno}

\documentclass[onecolumn,preprintnumbers,amsmath,amssymb]{revtex4}
\usepackage{graphicx}
\usepackage{bm}
\usepackage{epsfig}
\usepackage{subfigure}
\usepackage{xcolor}

\newcommand{\bi}{\begin{itemize}}
\newcommand{\ei}{\end{itemize}}
\newcommand{\be}{\begin{eqnarray}}
\newcommand{\ee}{\end{eqnarray}}
\newcommand{\beq}{\begin{equation}}
\newcommand{\eeq}{\end{equation}}
\newcommand{\beqn}{\begin{equation*}}
\newcommand{\eeqn}{\end{equation*}}
\newcommand{\bbmatrix}{\left( \begin{array}}
\newcommand{\eematrix}{\end{array} \right)}

\def\change#1{\textcolor{black}{#1}}
\def\dd{\text{d}}

\begin{document}

\title{Application of Optimal Control to Time-Resolution Protocol for Quantum Sensing }

\author{Chungwei Lin$^{1}$\footnote{clin@merl.com}, Qi Ding$^{2}$,  Yanting Ma$^{1}$}

\affiliation{Mitsubishi Electric Research Laboratories (MERL), 201 Broadway, Cambridge, MA 02139}

\affiliation{$^2$Department of Electrical Engineering and Computer Science, Massachusetts Institute of Technology, Cambridge, MA 02139, USA}



\date{\today}
\begin{abstract}
Time-resolution protocol of quantum sensing aims to measure the fast temporal variation of an external field and demands a high field sensitivity in a short interrogation time $\tau$.
Since any operation that evolves the quantum state takes time and is counted as part of the interrogation, evaluating the performance of time-resolution protocol requires a complete end-to-end description of the measurement process.
In particular, the initial state has to be one of the quantum sensor's eigenstates in the absence of external fields, and the final projective measurements must be performed in the same eigenstate basis.
Building upon prior works which proposed limits for time-resolved sensing using a quantum sensor, we apply optimal control theory to optimize the time-resolution protocol. Our analysis indicates that there exists a critical interrogation time $T^*$: when $\tau<T^*$ the optimal protocol is purely bang-bang; when $\tau>T^*$ the optimal protocol involves a singular control during the interrogation.
In the short-$\tau$ regime, which is relevant to high time resolution, we propose a ``detune protocol'' that involves only {\em smooth} control during the entire interrogation. As the discontinuities of control pose the main obstacles to experimental realization,
we expect the presented detune protocol to be practically useful.
In the long-$\tau$ regime, the optimal protocol closely resembles the Ramsey sequence; protocols based on maximizing Quantum Fisher Information are constructed to highlight the difference between the theoretically optimal and practically implementable measurements.
Effective use of the time-resolution protocol requires a setup where the unknown time-domain signal of interest can be {\em identically} and {\em repeatedly} generated.
As a potentially relevant application, we outline the calibration of baseband flux pulse distortion in the control of superconducting qubits.
\end{abstract}

\maketitle

\section{Introduction}

Quantum sensing generally refers to the technology that explicitly makes use of quantum phenomena to achieve highly sensitive and precise measurement \cite{RevModPhys.89.035002}. Notable examples include employing the squeezed state of light to enhance the sensitivity of laser interferometers \cite{PhysRevLett.59.278, PhysRevLett.59.2153, Breitenbach-1997, PhysRevLett.123.231107, PhysRevLett.123.231108},
using Nitrogen-Vacancy (NV) centers in diamond to measure signals generated by sub-micron objects \cite{RevModPhys.92.015004, Glenn-2018, Ku-2020, PhysRevApplied.14.014097}, and utilizing the multi-level interference spectroscopy of Rydberg atoms to determine the amplitude of microwave and terahertz electromagnetic signals \cite{RevModPhys.82.2313, Sedlacek-2012, PhysRevLett.111.063001, Fan-2014, 10.1063/1.5028357, 10.1063/1.5099036, PhysRevX.10.011027, PhysRevApplied.21.044025}. Intrinsic quantum advantage amounts to identifying and preparing the quantum state that is most sensitive to the field to be measured. The sensing performance can be quantified by the Quantum Fisher Information (QFI) \cite{book:Helstrom, book:Holevo}; given an unbiased estimator, the inverse of QFI gives the lower bound (Cram\'er-Rao bound) of the variance of the parameter to be estimated.
QFI, which depends solely on the quantum state and not on the specific measurement observables,
is very valuable in both formal analysis and numerical simulations. From a theoretical point of view, QFI provides a framework for comparing performances of different states. Given a sensing setup, analysis of QFI identifies its theoretical bound (known as the Heisenberg limit) and the corresponding best-sensing state (equal superposition of eigenstates with the largest and smallest eigenvalues)~\cite{PhysRevLett.96.010401, AdvancesQuantumMetrology_2011}.
From a practical perspective, maximizing QFI over the available control parameters offers a concrete means -- or at least a guide -- for approaching the quantum limit~\cite{PhysRevA.96.012117, PhysRevA.96.042114, PhysRevLett.124.060402, PhysRevA.103.052607, PhysRevA.105.042621}. 

For realistic implementations, both the initial state and the measurement observable have to be specified to determine the overall sensing performance; they are especially essential for the time-resolution protocols.
The objective of the time-resolution protocol is to maximize the sensitivity with respect to the to-be-measured field in a short interrogation time $\tau$. 
Because any operation that evolves the quantum state takes time,
the description of the time-resolution protocol should begin with the quantum sensor in one of its eigenstates in the absence of external fields, and end up with a projective measurement in the same eigenstate basis.
In Ref.~\cite{PhysRevLett.133.210802}, Herb and Degen propose a time-resolution protocol based on the Ramsey sequence \cite{PhysRev.78.695} and the Rotating Wave Approximation (RWA). Their protocol, which will be served as the reference for comparison, consists of  two sequential pulses of equal duration, one corresponding to $\hat{\sigma}_y$ and the other to $\hat{\sigma}_x$ in the rotating frame, and has been experimentally demonstrated \cite{Herb_2025}.

In this work, we apply Optimal Control Theory (OCT) to analyze the time-resolution protocol. OCT provides an efficient way to minimize a user-defined terminal cost subject to the dynamics that contains a time-dependent scalar control  \cite{book:Luenberger, book:Liberzon, book:GeometricOptimalControl, book:Pontryagin}.  It has emerged as a valuable tool for engineering a wide range of quantum systems \cite{doi:10.1116/5.0006785, PRXQuantum.2.030203, PRXQuantum.2.010101, ansel2024introduction}, such as stabilizing ultracold molecules \cite{PhysRevA.70.013402}, optimizing the performance in nuclear magnetic resonance measurement \cite{PhysRevA.90.023411, KOBZAR2012142, PhysRevA.101.012321}, cooling of quantum systems \cite{PhysRevA.82.063422, doi:10.1137/100818431, PhysRevA.87.043607}, \change{ and enhancing the yield of the target product in a chemical process \cite{PhysRevLett.104.220502, doi:10.1021/acs.jpclett.2c02840, PRXQuantum.5.020303, smith2025interradicalmotionpushmagnetosensing}}.
Apart from the better sensitivity, which is expected as the optimal control gives the best possible performance, the general behavior of the optimal protocols is qualitatively similar to that of the reference protocols.
In the short-$\tau$ regime which is relevant to high time resolution, we propose a detune protocol that only involves the {\em smooth} control field over the entire interrogation and performs comparably to the reference protocol.
Given that discontinuities in the control field pose the primary experimental challenge, we expect the detune protocol to be practically useful.
One bottleneck of meaningfully utilizing the time-resolution protocol is the need to generate the {\em identical}, yet unknown time-domain signal repeatably. We suggest that it can be used to calibrate the pulse distortion in fast flux control in superconducting qubits \cite{10.1063/1.5133894, hellings2025calibratingmagneticfluxcontrol,Li_2025}, which is an essential but cumbersome process for calibrating quantum gates, and is increasingly detrimental as the number of qubits gets larger. We discuss the potential to substantially reduce the total calibration time by leveraging our protocol to perform sequential measurements of the same distorted pulse. 
The rest of the paper is organized as follows. Section II defines the problem of time resolution and provides the necessary background. Section III presents our main results, including the general behavior from optimal control and a detailed description of the detune protocol. Section IV discusses practical aspects, including an outline of a realistic application. Section V provides a brief conclusion.

\section{Problem statement and optimal control}

\subsection{Overview and problem statement} \label{sec:overview}

In this subsection we formulate the design of the time-resolution protocol for quantum sensing as an optimal control problem. \change{ The quantum sensor is modeled as a two-level system, which can physically correspond to an electron spin, an atom, or a superconducting qubit; it is described by the Hamiltonian } \cite{PhysRevLett.133.210802}
\beq
H = \frac{\omega_0 + \delta \omega}{2} \hat{\sigma}_z + u(t) \hat{\sigma}_x.
\label{eqn:H_basic}
\eeq
where $\omega_0$ is the natural angular frequency of the quantum sensor, $\delta \omega$ is the field which we want to measure, and $u(t)$ is a scalar field which we can control. We shall take $\hbar \equiv 1$, $\omega_0 \equiv 1$ in our simulations but explicitly keep $\omega_0$ in all expressions; within this convention energy is measured in units of $\hbar \omega_0$ and time $\omega_0^{-1}$.  $| 0 \rangle$ and $|1 \rangle$ are eigenstates of $\hat{\sigma}_z$ with eigenvalues of 1 and -1 respectively. They are eigenstates without external fields (i.e., $\delta \omega = u = 0$) and will be referred to as the natural eigenstates of the quantum sensor.

The performance of the time-resolution protocol is quantified by its measurement sensitivity with respect to $\delta \omega$ given an (short) interrogation time $\tau$. Because any operation contributes to the interrogation time, an end-to-end description of the measurement process is required for performance evaluation. In particular the initial state and the basis of the final projection measurement have to be the sensor's natural eigenstates. Without loss of generality we choose the initial state to be $| 0 \rangle $ and measure the probability that ends up with the same $| 0 \rangle $ state; the outcome probability can be expressed as
\beq
\begin{aligned}
p (\delta \omega) & \equiv | \langle 0 | U[u; \delta \omega] | 0 \rangle |^2  \\
&\approx \underbrace{ | \langle 0 | U[u] | 0 \rangle |^2 }_{ \equiv p_0 }  + \underbrace{ (\delta \omega)  \frac{\partial}{\partial (\delta \omega) } | \langle 0 | U[u] | 0 \rangle |^2 }_{ \equiv \delta p } \\
&\approx p_0 + \eta \cdot (\delta \omega). 
\end{aligned}
\label{eqn:p_expansion_def}
\eeq
$p_0$ is the probability with zero field (i.e., $\delta \omega=0$) and is independent of $\delta \omega$; the linear coefficient of $\delta \omega$, denoted as $\eta$, is the ``measurement sensitivity'' \cite{PhysRevLett.133.210802}:
\beq
\eta = \frac{\partial}{\partial (\delta \omega) } | \langle 0 | U[u] | 0 \rangle |^2.
\label{eqn:eta_defintion}
\eeq
Larger $|\eta|$ implies a greater capability of detecting a small signal. For the rest of paper we take $\eta = |\eta|$ and neglect its sign. When $\delta \omega$ is small, the linear approximation is sufficient and the unbiased estimator of $\delta \omega$ is given by
\beq
\begin{aligned}
  \delta \omega = \frac{ p - p_0  }{ \eta  }.
\end{aligned}
\label{eqn:unbiased_estimator}
\eeq

In terms of optimization, the objective is to find the protocol $u(t)$ that maximizes the amplitude of sensitivity $|\eta|$ or $\eta^2$, given an interrogation time $\tau$ and an amplitude constraint $u_\text{max}$. We will apply OCT to obtain the optimal protocol and its corresponding maximum sensitivity.
Having identified the best possible performance, we introduce a smooth but sub-optimal  ``detune protocol'' that addresses experimental feasibility.
It is worth pointing out that the ultimate goal of time-resolution protocol is to reconstruct an unknown $\delta \omega(t)$ via $p( \delta \omega(t) )$, but to measure $p( \delta \omega(t) )$ requires repeatedly generating an identical yet unknown $\delta \omega(t)$. Identifying realistic settings that satisfy these conditions is also part of the problem, and
we shall consider a potentially meaningful application in Section \ref{sec:Dicussion_appl}.


\subsection{Reference: resonant YX protocol}

The reference protocol, which will be termed as the resonant YX protocol or simply the YX protocol, is the one proposed by Herb and Degen in Ref.~\cite{PhysRevLett.133.210802}; its control waveform is
\beq
u_\text{YX}(t) =
\begin{cases}
 u_\text{max} \cos( \omega_0 t + \frac{\pi}{2} ) & 0 \leq t \leq \text{min} (\frac{\tau}{2}, \frac{t_\text{QSL}}{2})  \\
0 & \text{min} (\frac{\tau}{2}, \frac{t_\text{QSL}}{2}) \leq t \leq \text{max}( \frac{\tau}{2}, \tau- \frac{t_\text{QSL}}{2} ) \\
 u_\text{max} \cos( \omega_0 t  ) &
 \text{max}( \frac{\tau}{2}, \tau- \frac{t_\text{QSL}}{2} )  \leq t \leq \tau 
\end{cases},
\label{eqn:Resonant_YX}
\eeq
where $t_\text{QSL} = \frac{\pi}{u_\text{max}}$ is referred to as the quantum speed limit \cite{PhysRevLett.133.210802, QuantumSpeedLimit, Mandelstam_1945, Norma_1998, Deffner_2017}. The rationale behind the protocol naming is as follows: the ``resonant'' indicates that the applied frequency is identical to the natural frequency; ``YX''  that in RWA, $\cos( \omega_0 t + \frac{\pi}{2} ) \hat{\sigma}_x$ and $\cos( \omega_0 t )\hat{\sigma}_x$ respectively lead to $\hat{\sigma}_y$ (Y operation) and $\hat{\sigma}_x$ (X operation) in the rotating frame. Overall the YX protocol identifies a critical interrogation time $t_\text{QSL}$, below which the protocol is composed of equal-duration Y and X operations, and above which the protocol involves a free evolution between Y and X operations.

With RWA, the sensitivity of the YX protocol grants an analytical expression
\beq
\begin{aligned}
\eta_\text{YX,RWA}(\tau) 
&= \begin{cases} \frac{ t_\text{QSL} }{\pi}
  \sin(  \frac{\pi}{2}  \frac{ \tau }{ t_\text{QSL} }  )  \big( 1 - \cos(\frac{\pi}{2}  \frac{ \tau }{ t_\text{QSL} }) \big) & \tau < t_\text{QSL} \\
  \frac{t_\text{QSL}}{2} \big( \frac{\tau}{t_\text{QSL}} - (1 - \frac{2}{\pi }) \big)
 & \tau > t_\text{QSL}
\end{cases}.
\end{aligned}
\label{eqn:eta_YX_RWA_full_time}
\eeq
Since $\eta$ has the dimension of time, it is suitable to introduce a dimensionless measure $\frac{\eta}{\tau}$. When expressed in the dimensionless time variable $\tilde{\tau} \equiv \frac{\tau}{t_\text{QSL}}$, $\frac{\eta}{\tau}$ is given by
\beq
\frac{\eta_\text{YX,RWA}}{\tau} = \begin{cases} \frac{ \tilde{\tau}^{-1} }{\pi}
  \sin(  \frac{\pi}{2}  \tilde{\tau}  )  \big( 1 - \cos(\frac{\pi}{2}  \tilde{\tau} ) \big) & \tilde{\tau}<1 \\
  \frac{ \tilde{\tau}^{-1} }{2} \big( \tilde{\tau} - (1 - \frac{2}{\pi }) \big)
 & \tilde{\tau} > 1
 \end{cases}.
 \label{eqn:eta/tau_YX_RWA_full_time}
\eeq
Eq.~\eqref{eqn:eta/tau_YX_RWA_full_time} has no explicit $u_\text{max}$ dependence and approaches $\frac{1}{2}$ as $\tilde{\tau} \rightarrow \infty$. When comparing performances of different $u_\text{max}$'s, $\tau$'s, and protocols, it is convenient to plot $\frac{\eta}{\tau}$ as a function of $\frac{\tau}{t_\text{QSL}}$; this convention will be adopted in this work and Eq.~\eqref{eqn:eta/tau_YX_RWA_full_time} will be served as the baseline.

When designing the protocol we assume that $\delta \omega$ stays unchanged during the entire (short) interrogation $\tau$, thus within this setting $\tau$ is the minimum time that a time-resolution protocol can resolve. Once the protocol is fixed, however, a more general expression that includes the fast time variation of $\delta \omega$ is given by
\beq
\delta p(\delta \omega) = \int_{t - \tau/2}^{t+\tau/2} \dd t' K(t'-t) \cdot \delta \omega(t').
\label{eqn:convolution}
\eeq
Here $K$ is a protocol dependent convolution kernel; for constant $\delta \omega$ one recovers $\eta = \int_{-\tau/2}^{\tau/2} \dd t K(t)$. In Appendix \ref{app:kernel} (see also Supplementary of Ref.~\cite{PhysRevLett.133.210802}) we provide a simple numerical method to construct the kernel of Eq.~\eqref{eqn:convolution}. Once $K$ is known, the time resolution shorter than $\tau$ is in principle possible \cite{Herb_2025}.

To facilitate the subsequent discussion we explain our subscript naming convention. The sensitivity $\eta$ can be evaluated using different protocols and approximations; they are indicated in the subscript by $\eta_\text{ Protocol, Approximation}$. In this work, 'Protocol' can be one of (i) YX (reference), (ii) D (detune), or (iii) Opt (optimal); 'Approximation' can be either (i) RWA or (ii) full calculation (no approximation). We neglect the second subscript for full calculation. For example, $\eta_{D}$ is obtained using the detune protocol and full calculation.

\subsection{Optimal control} \label{sec:optimal_control}

In this subsection we summarize the important functions introduced in OCT. Numerous comprehensive reviews have been reported in the literature \cite{PRXQuantum.2.030203, PRXQuantum.2.010101}, and its applications specific to the unitary qubit can be found in Ref.~\cite{PhysRevA.111.042602}. Here, special attention is devoted to the cost function that involves a derivative of the wave function.

The sensitivity $\eta$ characterizes the sensing performance and is the most important quantity to evaluate. Its explicit dependence on the final-time wave function $| \psi_0 (\tau) \rangle$ is given by
\beq
\begin{aligned}
\eta &= \frac{\partial}{\partial (\delta \omega) }
\bigg[ \langle 0 | \psi_0 (\tau) \rangle \langle \psi_0 (\tau) | 0 \rangle \bigg]
 \\
&= \langle 0 | \psi_1 (\tau) \rangle \langle \psi_0(\tau) | 0 \rangle +
\langle 0 | \psi_0 (\tau) \rangle \langle \psi_1 (\tau) | 0 \rangle,  \\
\text{ where } &
| \psi_1 (t) \rangle \equiv  \frac{\partial}{\partial (\delta \omega) } | \psi_0 (t)  \rangle.
\end{aligned}
\label{eqn:eta_evaluation}
\eeq
Optimizing the time-resolution protocol is equivalent to maximizing $|\eta|$. To avoid dealing with sign of $\eta$ we choose to maximize $\eta^2$ or equivalently to minimize the terminal cost function
\beq
\begin{aligned}
\mathcal{C}_{\eta^2} &= -\frac{1}{2} \eta^2
=-\frac{1}{2} \left\{ \frac{\partial}{\partial (\delta \omega) }
\big[ \langle 0 | \psi_0 (\tau) \rangle \langle \psi_0 (\tau) | 0 \rangle \big]
\right\}^2 \\
&= -\frac{1}{2} \big( \langle 0 | \psi_1 (\tau) \rangle \langle \psi_0(\tau) | 0 \rangle +
\langle 0 | \psi_0 (\tau) \rangle \langle \psi_1 (\tau) | 0 \rangle \big)^2.  
\end{aligned}
\label{eqn:cost_time_resolution}
\eeq
The optimization problem can now be formally specified:
\beq
\begin{aligned}
&\text{Find $u^*(t)$ that minimizes } \mathcal{C}_{\eta^2} [u(t)] \text{  subject to} \\
& \text{(i) dynamics }   i \partial_t | \psi_0 (t) \rangle = \big[ \frac{\omega_0 + \delta \omega}{2} \hat{\sigma}_z + u(t) \hat{\sigma}_x \big] | \psi_0(t) \rangle, \\
&\text{(ii) initial state } | \psi(0) \rangle = |0\rangle, \\
&\text{(iii) total interrogation time $\tau$,} \\
&\text{(vi) amplitude constraint $|u(t)| \leq u_\text{max}$}.
\end{aligned}
\label{eqn:Opt_problem}
\eeq

OCT is a powerful tool for dynamics-constrained optimization. Given the dynamics and the cost to minimize, it introduces the adjoint variables $|\pi(t) \rangle$, the control-Hamiltonian $\mathcal{H}_\text{oc}(t)$, and the switching function $\Phi(t)$ [defined shortly in Eqs.~\eqref{eqn:OC_original_gradient}] that are computationally accessible and accurately
characterize the system response to some external or model variables. One non-trivial complication from the cost $\mathcal{C}_{\eta^2}$ is its dependence on  $| \psi_1 (\tau) \rangle =  \frac{\partial}{\partial (\delta \omega) } | \psi_0 (\tau)  \rangle$. To cope with this additional dependence we augment the system by regarding $| \psi_0 \rangle$ and $| \psi_1  \rangle$ as {\em independent} variables; more details can be found in Ref.~\cite{PhysRevA.103.052607, PhysRevA.101.022320}.

Denote the variables in augmented dynamics as $| \psi \rangle = \begin{bmatrix} | \psi_0 \rangle \\ | \psi_1 \rangle \end{bmatrix} $ and its adjoint variables as $| \pi \rangle = \begin{bmatrix} | \pi_0 \rangle \\ | \pi_1 \rangle \end{bmatrix} $, the OCT for a general terminal cost $\mathcal{C}( |\psi_0 (\tau) \rangle, |\psi_1 (\tau) \rangle )$ is summarized as follows.
\begin{subequations}
\begin{align}
\mathcal{H}_\text{oc}(t) &= \text{Im} \left\{
\begin{bmatrix} \langle \pi_0 | & \langle \pi_1 | \end{bmatrix}
\begin{bmatrix} H_0 & 0 \\ \frac{\hat{\sigma}_z}{2} & H_0 \end{bmatrix}
\begin{bmatrix} | \psi_0 \rangle \\ | \psi_1 \rangle \end{bmatrix}
\right\} \sim \frac{\partial \mathcal{C} }{\partial \tau},
\label{eqn:H_oc_0} \\
\Phi(t) &= \text{Im} \left\{
\begin{bmatrix} \langle \pi_0 | & \langle \pi_1 | \end{bmatrix}
\begin{bmatrix} \hat{\sigma}_x & 0 \\ 0 & \hat{\sigma}_x \end{bmatrix}
\begin{bmatrix} | \psi_0 \rangle \\ | \psi_1 \rangle \end{bmatrix}
\right\} = \frac{1}{\dd t} \frac{\delta \mathcal{C}}{\delta u(t) },
\label{eqn:Phi} \\
 \partial_t \begin{bmatrix} | \psi_0 \rangle \\ | \psi_1  \rangle \end{bmatrix} &= -i \begin{bmatrix} H_0 & 0 \\ \frac{\hat{\sigma}_z}{2}  & H_0  \end{bmatrix}
 \begin{bmatrix} | \psi_0 \rangle \\ | \psi_1   \rangle \end{bmatrix}
 \text{ with } \begin{bmatrix} | \psi_0(0) \rangle \\ | \psi_1(0)  \rangle \end{bmatrix}= \begin{bmatrix} | 0 \rangle \\ 0  \end{bmatrix},
 \label{eqn:forward}  \\
 \partial_t \begin{bmatrix} | \pi_0 \rangle \\ | \pi_1  \rangle \end{bmatrix} &
= -i
 \begin{bmatrix} H_0 & \frac{\hat{\sigma}_z}{2} \\ 0 & H_0  \end{bmatrix}
 \begin{bmatrix} | \pi_0 \rangle \\ | \pi_1   \rangle \end{bmatrix} \text{ with } \begin{bmatrix} | \pi_0(\tau) \rangle \\ | \pi_1(\tau)  \rangle \end{bmatrix}= 2 \begin{bmatrix} \frac{\partial \mathcal{C}}{\partial \langle \psi_0(\tau)| } \\ \frac{\partial \mathcal{C}}{\partial \langle \psi_1(\tau)| } \end{bmatrix},
 \label{eqn:adjoint}  
\end{align}
\label{eqn:OC_original_gradient}
\end{subequations}
where  $H_0 = H(\delta \omega=0) = \frac{\omega_0}{2} \hat{\sigma}_z + u(t) \hat{\sigma}_x$. Terminal cost determines the final-time condition of the adjoint variables $| \pi \rangle$. Two terminal costs will be considered. To maximize the sensitivity we choose $\mathcal{C} = \mathcal{C}_{\eta^2}$ so that
\beq
\begin{aligned}
\begin{bmatrix} | \pi_0(\tau) \rangle \\ | \pi_1(\tau)  \rangle \end{bmatrix} &= 2 \begin{bmatrix} \frac{\partial \mathcal{C}_{\eta^2} }{\partial \langle \psi_0(\tau)| } \\ \frac{\partial \mathcal{C}_{\eta^2} }{\partial \langle \psi_1(\tau)| } \end{bmatrix}
= - 4 \, \text{Re} \big[ \langle 0 | \psi_0 (T) \rangle \langle \psi_1 (T) | 0 \rangle  \big]
\begin{bmatrix}
| 0 \rangle  \langle 0 | \psi_1 (\tau) \rangle  \\
 | 0 \rangle  \langle 0 | \psi_0 (\tau) \rangle
\end{bmatrix}.
\end{aligned}
\label{eqn:pi_boundary_square}
\eeq
We also consider maximizing QFI \cite{PhysRevA.103.052607} where the terminal cost is negative QFI:
\beq
\begin{aligned}
\mathcal{C}_\text{QFI} &= - \text{QFI} = - 4 \left[ \langle \psi_1(\tau) | \psi_1(\tau) \rangle - |  \langle \psi_0(\tau) | \psi_1(\tau) \rangle |^2 \right], \\
\begin{bmatrix} | \pi_0(\tau) \rangle \\ | \pi_1(\tau)  \rangle \end{bmatrix}
&= 2 \begin{bmatrix} \frac{\partial \mathcal{C}_\text{QFI} }{\partial \langle \psi_0(\tau) |} \\ \frac{\partial \mathcal{C}_\text{QFI} }{\partial \langle \psi_1(\tau) |}  \rangle \end{bmatrix}
= 8 \begin{bmatrix}
  | \psi_1 (\tau) \rangle  \langle \psi_1 (\tau) | \psi_0 (\tau) \rangle, \\
 -  | \psi_1(\tau) \rangle +  | \psi_0 (\tau) \rangle  \langle \psi_0 (\tau) | \psi_1(\tau) \rangle
\end{bmatrix}.
\end{aligned}
\label{eqn:pi_boundary_QFI}
\eeq
In Section \ref{sec:long_tau} we present the results from protocols that maximizes QFI.

Eqs.~\eqref{eqn:OC_original_gradient} plus Eq.~\eqref{eqn:pi_boundary_square} [or Eq.~\eqref{eqn:pi_boundary_QFI}] completely determine quantities introduced by OCT. They are used to express two general optimality conditions \cite{control_type}. First, the optimal control is
\beq
u^*(t) = \begin{cases}
          -u_\text{max} \text{Sgn}[\Phi] & \text{ if } \Phi \neq 0  \text{, bang (B) control } \\
          u^\text{sing}(t) &\text{  if } \Phi = 0 \text{, singular (S) control }
         \end{cases},
\label{eqn:u^*(t)}
\eeq
where Sgn denotes the sign function.
The values of singular control $u^\text{sing}$ need to be determined numerically, but OCT provides the following expression [see Appendix \ref{App:singular_value} for the derivation]
\beq
\begin{aligned}
  u^\text{sing} (t)   &= \frac{\omega_0}{2}
  \frac{ \text{Im} [  \langle \pi | \begin{bmatrix} 0&0 \\ \hat{\sigma}_x & 0 \end{bmatrix}   | \psi \rangle ]  }{ \text{Im} [  \langle \pi |
  \begin{bmatrix} \frac{\omega_0}{2} \hat{\sigma}_z &0 \\ \frac{1}{2} \hat{\sigma}_z & \frac{\omega_0}{2} \hat{\sigma}_z \end{bmatrix}
   | \psi \rangle ] },
  \end{aligned}
\label{eqn:singular_control_value}
\eeq
which is very valuable for numerics.
Second, for an optimal solution,  $\mathcal{H}_\text{oc}(t)$ is a constant over the entire evolution and its value is proportional to the derivative of the terminal cost function with respect to the interrogation time $\tau$. For both $\mathcal{C}_{\eta^2}$ and $\mathcal{C}_\text{QFI}$, the resulting $\mathcal{H}_\text{oc}$'s are always negative as increasing $\tau$ always increases $\eta^2(\tau)$ and QFI$(\tau)$.

We emphasize that although the physical system considered here is equivalent to a qubit with unitary dynamics [i.e., Eq.~\eqref{eqn:H_basic}], the augmented dynamics specified in Eq.~\eqref{eqn:forward} is neither planar (i.e., two real-valued variables) nor unitary. For these reasons certain properties specific to the unitary qubit  given in Ref.~\cite{PhysRevA.111.042602} do not hold anymore. Particularly we point out that the BB control does {\em not} have identical bang duration and the singular control does {\em not} require $u^\text{sing}=0$.

\subsection{Practical use of OCT functions}

We conclude this section by highlighting the usefulness of OCT. The most practically useful quantity is undoubtedly the switching function $\Phi$. By sequentially solving the dynamics twice, one forward in time for $| \psi \rangle$ and one backward  $| \pi \rangle$, one is able to obtain the gradient of $\mathcal{C}$ with respect to all $u(t_i)$ \cite{KHANEJA2005296}. Three applications used in the work are highlighted. First $\Phi$ is primarily used to construct a numerical $u^*(t)$ by some gradient-based (or quasi-Newton) algorithm that involves
\beq
u^{(n+1)}(t) \rightarrow u^{(n)}(t) - \text{updating rate} \times \Phi.
\label{eqn:update_gradient}
\eeq
For the free-form parametrization one divides $\tau$ into $N_t$ intervals and assumes a piecewise constant waveform. In our simulations $N_t$ ranges from 100 to 2000 (depending on $\tau$); we ensure that the result has a negligible change upon increasing $N_t$'s and that $N_t$ is sufficiently large so that all optimality conditions are numerically satisfied \cite{PhysRevA.103.052607}. Empirically we find that when $u^*(t)$ contains both B and S controls, the convergence of S control is slow due to its vanishing gradient. In this case we can compute Eq.~\eqref{eqn:singular_control_value} at $n$th iteration, and replace the non-bang controls by the singular values:
\beq
u^{(n+1)}(t) \rightarrow u^\text{sing,$(n)$} (t) \text{ for } | u^{(n)}(t) | < u_\text{max}.
\label{eqn:update_u_sing}
\eeq
Including Eq.~\eqref{eqn:update_u_sing} greatly accelerates the convergence.

Second, quite often the feasible control waveform is constrained by the hardware and is conveniently parametrized by $u(t; \vec{\alpha})$;
the objective in this case reduces to optimizing the performance within the parametrized control space.  The gradient $\frac{\partial \mathcal{C}}{\partial \vec{\alpha}}$  can be computed by
\beq
\frac{\partial \mathcal{C} }{ \partial \vec{\alpha} } = \sum_t
\frac{\partial \mathcal{C} }{ \partial u(t) } \frac{\partial u(t; \vec{\alpha})}{\partial \vec{\alpha} } = \sum_t \dd t\, \Phi(t)  \frac{\partial u(t; \vec{\alpha})}{\partial \vec{\alpha} } = \int_0^\tau \dd t \,\Phi(t)  \frac{\partial u(t; \vec{\alpha})}{\partial \vec{\alpha} }.
\label{eqn:gradeint_alpha}
\eeq
One then uses some gradient-based algorithm to find the solution. Eq.~\eqref{eqn:gradeint_alpha} will be used to construct the best detune protocol [see Section \ref{sec:detune_opt}].

Third, $\Phi$ can be used to estimate the deviation in performance caused by the imperfect control. For this application, one computes $\Phi_\text{ref}(t)$ using the reference control $u_\text{ref}(t)$. If the applied control $u_\text{app}(t)$ is different from the reference control, the resulting change in cost can be estimated by
\beq
\mathcal{C}[u(t)] - \mathcal{C}[u_\text{ref}(t)]
= \sum_i \frac{ \partial \mathcal{C} }{ \partial u_i } [ u_\text{app}(t_i) - u_\text{ref}(t_i) ]
= \int_0^\tau \dd t \, \Phi_\text{ref} (t ) [ u_\text{app}(t) - u_\text{ref}(t) ].
\label{eqn:deviation_control}
\eeq
Eq.~\eqref{eqn:deviation_control} can be used to estimate the performance reduction due to non-ideal switching times in BB protocol. It will be used to estimate the effect of smoothness (or uncertainty of exact switching times in BB protocols) in Section \ref{sec:smoothness}.

Two practical utilizations of control-Hamiltonian $\mathcal{H}_\text{oc}(t) \sim \frac{\partial \mathcal{C}}{\partial \tau}$ are outlined. First, the deviation from a constant  $\mathcal{H}_\text{oc}(t)$ can be used to quantify the solution quality \cite{PhysRevA.103.052607}. Second, in a damped quantum system $\mathcal{H}_\text{oc} = 0$ offers a rigorous criterion for defining the optimal operation time for a quantum task \cite{PhysRevA.105.042621}.

\section{Optimal protocol and realistic protocol}

\begin{figure}[ht]
\begin{center}
\includegraphics[width=0.98\textwidth]{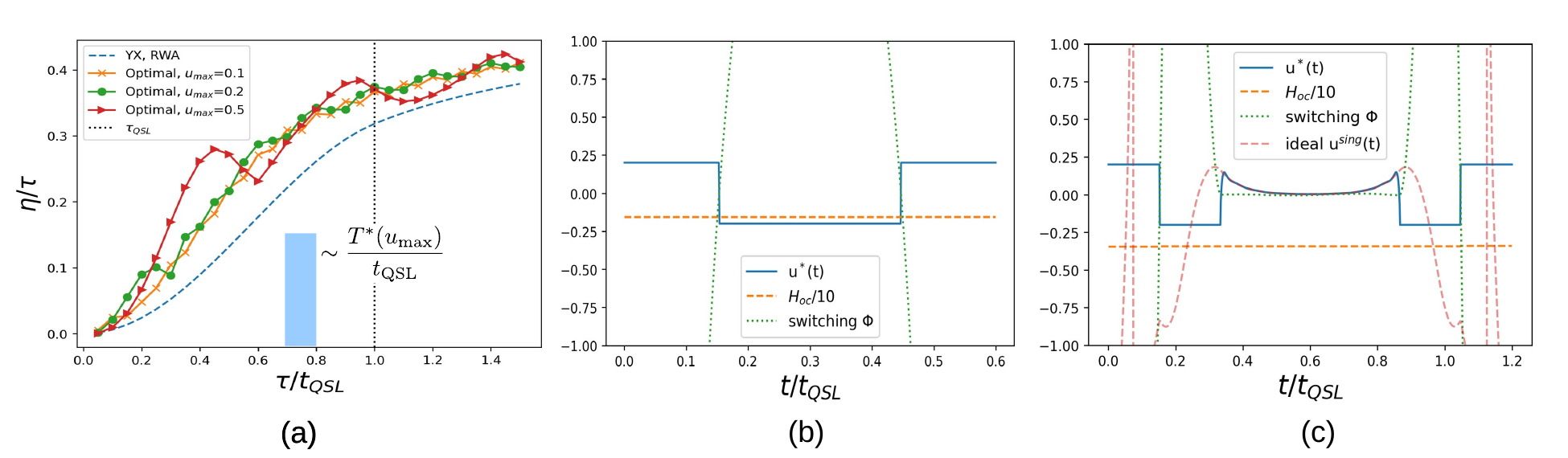}
\caption{
$\eta_\text{opt}/\tau$ for $u_\text{max}=0.1$, 0.2, 0.5. The reference YX protocol using RWA is provided as the reference. The metrological gain of optimal protocol is more significant for short interrogation time, which is the regime relevant to high time resolution.  The critical interrogation time $T^*$ depends on $u_\text{max}$ and is about 0.8 $t_\text{QSL}$ ($t_\text{QSL} = \frac{\pi}{u_\text{max}}$) for $u_\text{max}<0.2$ (blue area). (b) For $u_\text{max}=0.2$, ${\tau} = 0.6 \, t_\text{QSL}$, the optimal control is of BB. Optimality conditions are verified. (c) For $u_\text{max}=0.2$, ${\tau} = 1.2 \, t_\text{QSL}$, the optimal control includes a singular portion in the middle.  Optimality conditions -- Eq.~\eqref{eqn:u^*(t)} for B control,  Eq.~\eqref{eqn:singular_control_value} for S control, and a constant $\mathcal{H}_\text{oc}(t)$ -- are numerically verified.
}
\label{fig:detune_protocol}
 \end{center}
\end{figure}

\subsection{Overview and general behavior}

This section contains our main results which include the performance of the optimal protocol and the detune protocol: the former represents the best sensitivity the system can possibly achieve; the latter is a {\em smooth} protocol with performance comparable to that of the reference YX protocol.

The overall behavior of optimal protocols is illustrated in Fig.~\ref{fig:detune_protocol}(a) where the optimal $\frac{\eta}{\tau}$ for $u_\text{max} = $0.1, 0.2, 0.5 and the reference YX protocol [using RWA, see Eq.~\eqref{eqn:eta/tau_YX_RWA_full_time}] are plotted. It is seen that the metrological gain of optimal control is more significant for the short interrogation time. Similar to the YX protocol, the optimal protocol introduces a critical interrogation time $T^*$. When $\tau<T^*$ the optimal protocol is exclusively BB; when  $\tau>T^*$ the optimal protocol involves a singular control in the middle, resembling the free evolution in the Ramsey protocol.
Two representative control waveforms are illustrated in Fig.~\ref{fig:detune_protocol}(b) and (c) where the optimal protocols for $\tau = 0.6 \, t_\text{QSL}$ and $1.2 \, t_\text{QSL}$ are plotted. The optimality conditions are numerically verified in both cases.
For the rest of this section, we provide more detailed analysis for both short ($\tau < T^*$) and long ($\tau > T^*$) interrogation times. 

\subsection{Short interrogation time $\tau < T^*$ I: detune protocol in RWA}

One practical challenge in implementing the reference YX protocol is its discontinuity at the midpoint of the interrogation. In this subsection we present a {\em smooth} detune protocol and show that within RWA it can achieve performance comparable to that of the YX protocol. The detune protocol is given by
\beq
u_\text{D,RWA}(t) = u_\text{max} \cos( \omega t + \theta ), \,\, 0 \leq t \leq \tau.
\label{eqn:detune_protocol_basic}
\eeq
The ``detune'' refers to the fact that the driving frequency $\omega/(2\pi)$ is different from the natural frequency $\omega_0/(2\pi)$; its value will be determined by maximizing the sensitivity $\eta$.

Applying the transformation  $| \tilde{\psi}_0 \rangle =  e^{-i \frac{\hat{\sigma}_z}{2} \omega t} | {\psi}_0 \rangle $ and neglecting the high-frequency terms, the resulting Schr\"odinger equation in the rotating frame is $i \partial_t | \tilde{\psi}_0 \rangle = H_\text{RWA}  | \tilde{\psi}_0 \rangle$ where the time-independent $H_\text{RWA}$ is
\beq
\begin{aligned}
H_\text{RWA} &= \frac{\delta \omega - \Delta \omega }{2} \hat{\sigma}_z + \frac{u_\text{max} }{2} \big( \hat{\sigma}_x \cos \theta + \hat{\sigma}_y \sin \theta \big) \\
& \equiv \frac{ \tilde{\Omega} }{2} \bigg[
\frac{ \Delta }{ \tilde{\Omega} } \hat{\sigma}_z + \frac{u_\text{max}}{ \tilde{\Omega} } \big( \hat{\sigma}_x \cos \theta + \hat{\sigma}_y \sin \theta \big)
\bigg].
\end{aligned}
\label{eqn:H_rot_0_rep}
\eeq
In Eq.~\eqref{eqn:H_rot_0_rep} $\Delta \equiv \delta \omega - \Delta \omega = \delta \omega - ( \omega - \omega_0)$ and $\tilde{\Omega}^2 = u_\text{max}^2 + \Delta^2$; $\Delta \omega = \omega - \omega_0$ is the detune of the applied angular frequency. 
The evolution in the rotating frame and its first diagonal component are
\beq
\begin{aligned}
 & e^{-i H_\text{RWA} \tau} =
 \cos( \frac{\tilde{\Omega} \tau }{2} ) \, \hat{e}_{2\times2}  -i \sin( \frac{\tilde{\Omega} \tau }{2} ) \bigg[ \frac{ \Delta }{ \tilde{\Omega} } \hat{\sigma}_z + \frac{u_\text{max}}{ \tilde{\Omega} } \big( \hat{\sigma}_x \cos \theta + \hat{\sigma}_y \sin \theta \big) \bigg] \\
  \Rightarrow & \big| \langle 0 | e^{-i H_\text{RWA} \tau} | 0 \rangle \big|^2 =
 \cos^2( \frac{\tilde{\Omega} \tau }{2} ) + \sin^2( \frac{\tilde{\Omega} \tau }{2} ) \cdot \frac{ \Delta^2 }{ u_\text{max}^2 + \Delta^2 }.
\end{aligned}
\label{eqn:evolution_X}
\eeq
Eq.~\eqref{eqn:evolution_X} shows that the phase $\theta$ is irrelevant as neither $\hat{\sigma}_x$ nor $\hat{\sigma}_y$ has diagonal components; this is not the case when using full calculation (next subsection). It is worth noting that $\big| \langle 0 | e^{-i H_\text{RWA} \tau} | 0 \rangle \big|^2 = \big| \langle 1 | e^{-i H_\text{RWA} \tau} | 1 \rangle \big|^2$, indicating the same protocol can apply to either natural eigenstate.

\begin{figure}[ht]
\begin{center}
\includegraphics[width=0.9\textwidth]{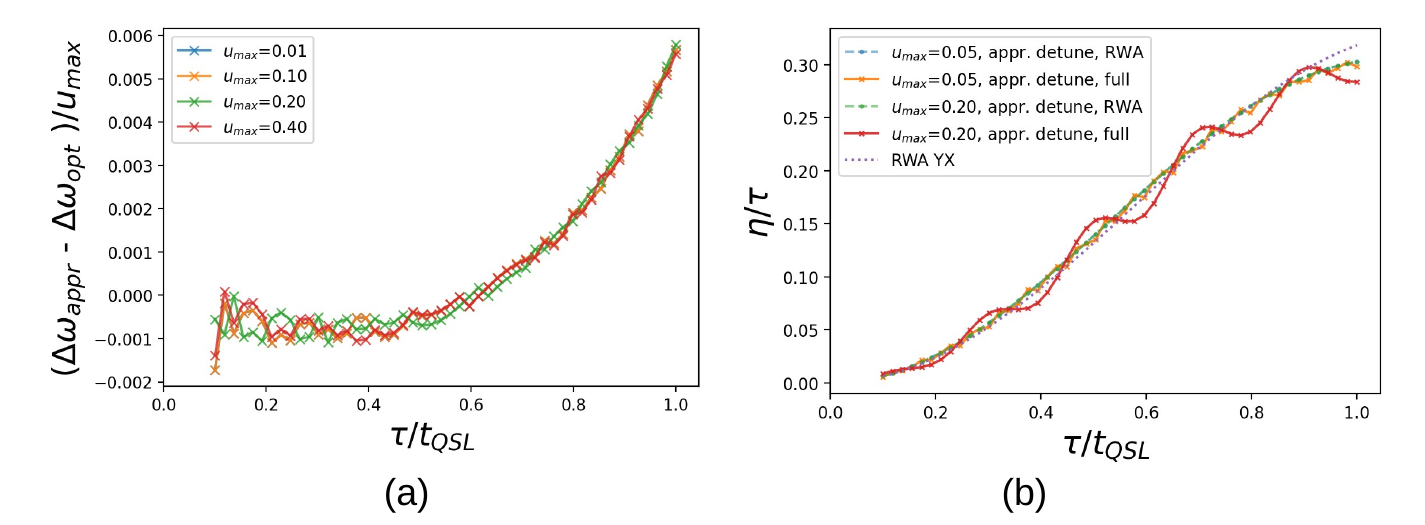}
\caption{Detune protocol within RWA.
(a) The difference of detune values between directly optimizing Eq.~\eqref{eqn:eta_Detune} and the approximation of \eqref{eqn:optimal_detune_RWA_approx}.  (b) The resulting $\eta/\tau$ for $u_\text{max}=0.05$ and 0.2. The difference between YX and detune protocols (RWA) is very small. The difference between detune protocols using RWA and the full calculation increases [using the approximate detune of \eqref{eqn:optimal_detune_RWA_approx}] as $u_\text{max}$ increases. 
}
\label{fig:detune_protocol_RWA}
 \end{center}
\end{figure}

The sensitivity of the detune protocol in RWA, denoted as $\eta_\text{D,RWA}$, is given by
\beq
\begin{aligned}
 \eta_\text{D,RWA} (\Delta \omega) &= \bigg[ \frac{\partial}{\partial (\delta \omega)} \big| \langle 0 | e^{-i H_\text{RWA} \tau} | 0 \rangle \big|^2 \bigg]_{ \delta \omega = 0}  \\
&= (\Delta \omega) \frac{  u_\text{max}^2  }{  \bar{\Omega}^4 }  \times \left[ \frac{ \bar{\Omega} \tau}{2} \sin( \bar{\Omega} \tau )
+ \cos(  \bar{\Omega} \tau  ) - 1 \right],
\end{aligned}
\label{eqn:eta_Detune}
\eeq
with  $\bar{\Omega}^2 = u_\text{max}^2 + (\Delta \omega)^2$.
$\eta_\text{D,RWA}$ being an odd function in $\Delta \omega$ indicates that a minimum at $\eta_\text{D,RWA} (\Delta \omega)$ guarantees a maximum at $\eta_\text{D,RWA}(-\Delta \omega)$ with the same amplitude $|\eta_\text{D,RWA}(\Delta \omega)| = |\eta_\text{D,RWA}(-\Delta \omega)|$. One bears in mind that the RWA more accurately describes the blue-detuned regime ($\Delta \omega > 0$) as the neglect of high-frequency components is better justified. Introducing the dimensionless parameter $x \equiv \bar{\Omega}\tau = \sqrt{ (u_\text{max} \tau)^2 + (\Delta \omega \cdot \tau)^2 } \geq u_\text{max} \tau$, Eq.~\eqref{eqn:eta_Detune} becomes
\beq
\begin{aligned}
 \frac{ \big| \eta_\text{D,RWA}(x) \big|  }{\tau}
 &=   (u_\text{max} \tau)^2  \sqrt{ x^2 - (u_\text{max} \tau)^2 } \, F(x) \text{ where } \\
 F(x) &= \frac{1}{x^4} \left[ 1 - \frac{ x }{2} \sin( x ) - \cos( x ) \right].
\end{aligned}
\label{eqn:eta_Detune_2}
\eeq
$F(x)$ is smooth at $x=0^+$ and positive for $0<x<2 \pi$.
The largest interrogation time we consider is $t_\text{QSL}$. The optimal detune protocol within RWA is to find $x$ and thus $\Delta \omega$ that maximizes $|\eta_\text{D,RWA}|$; the  optimized detune value depends on {\em both} $u_\text{max}$ and $\tau$. This can be done numerically and in the following we provide an analytical but accurate expression.

As a first-order approximation we assume the argmax $\eta_\text{D,RWA}(x) \gg u_\text{max} \tau$  so that $\sqrt{ x^2 - (u_\text{max} \tau)^2 } \approx x$. We can approximately maximize $\eta_0(x ) =  x F(x)$ and find $\text{argmax}_x \eta_0(x) = D_0 = 2.606$. With the same order of approximation, $D_0 = \Delta \omega \tau$ and thus $\Delta \omega \approx \frac{D_0}{\tau}$. This expression turns out to be good over for the entire $0< \frac{\tau}{ t_\text{QSL} } \leq 1$.
We numerically identify a small correction to this expression and the approximate detune is found to be
\beq
\Delta \omega_\text{appr} ( u_\text{max}, \tau ) \approx \frac{D_0}{\tau}
- 0.02 \cdot u_\text{max}^2 \tau
\label{eqn:optimal_detune_RWA_approx}
\eeq
The weak dependence on the $u_\text{max}$ reflects the nature of RWA.
The validity of Eq.~\eqref{eqn:optimal_detune_RWA_approx} is examined by comparing $\Delta \omega_\text{appr}$ with the optimal detune $\Delta \omega_\text{opt}$ obtained by directly maximizing Eq.~\eqref{eqn:eta_Detune}.
Fig.~\ref{fig:detune_protocol_RWA}(a) shows the difference between $\Delta \omega_\text{appr}$ and $\Delta \omega_\text{opt}$ is indeed small;  
Fig.~\ref{fig:detune_protocol_RWA}(b) shows the resulting sensitivities are very close to those of YX protocols. To quantify the error caused by RWA we also compute the detune protocol using the full calculation. The discrepancy between RWA and full calculation becomes more significant upon increasing $u_\text{max}$; for $\frac{u_\text{max}}{ \omega_0 } < 0.2$, their difference is smaller than 15\% when $\tau > 0.5 \, t_\text{QSL}$.

\begin{figure}[ht]
\begin{center}
\includegraphics[width=0.7\textwidth]{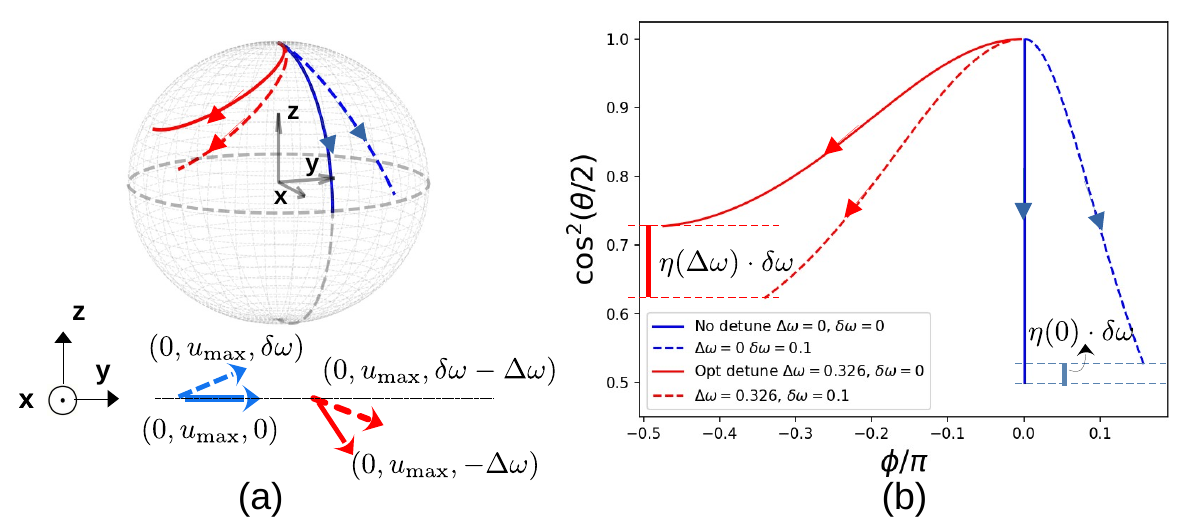}
\caption{Detune protocol in RWA and rotating frame. Trajectories on the Bloch sphere (a) and in $\phi$-$\cos^2 (\theta/2)$ plane (b). In both (a) and (b), the solid blue trajectory corresponds to the no-detune $\Delta \omega = 0$ and zero-field $\delta \omega = 0$ case; dashed blue to no-detune $\Delta \omega = 0$ and finite-field $\delta \omega = 0.1$; solid red to optimal-detune $\Delta \omega = 0.326$ and zero-field $\delta \omega = 0$; dashed red optimal-detune $\Delta \omega = 0.326$ and finite-field $\delta \omega = 0.1$. Within RWA, the detune protocol has a time-independent Hamiltonian and each trajectory corresponds to a rotating axis of the Bloch sphere. Bottom of (a) gives the rotating axes of four trajectories.
The $\cos^2(\theta/2)$ is the measurement signal, and the difference at the end of interrogation with and without field quantifies the sensing performance.
(b) In $\phi$-$\cos^2 (\theta/2)$ plane, the optimal detune (red) leads to a larger difference compared to the case of no detune (blue).
The trajectories are computed using $u_\text{max}=0.2$, $\tau = 0.5 \,t_\text{QSL} = 2.5 \pi$. 
}
\label{fig:detune_protocol_RWA_picture}
 \end{center}
\end{figure}

To provide some intuition, in Fig.~\ref{fig:detune_protocol_RWA_picture} we visualize how the detune protocol enhances sensitivity by comparing trajectories of two detune protocols on the Bloch sphere where the sensor state in rotating frame is parametrized by $| \tilde{\psi}_0 \rangle = \big[ \cos \frac{\theta}{2}, \sin \frac{\theta}{2} e^{i \phi} \big]^T$.
Within RWA, the detune protocol has a time-independent Hamiltonian and each trajectory is associated with a rotating axis of the Bloch sphere. Using Eq.~\eqref{eqn:H_rot_0_rep}, given a detune $\Delta \omega$ and a to-be-measured field $\delta \omega$, the rotating axis is pointing along $(0, u_\text{max}, \delta \omega - \Delta \omega)$. The square of projection $\cos^2 (\frac{\theta}{2})$ at the terminal point of the trajectory is the measurement signal.
As a concrete example we consider $u_\text{max}=0.2$ and $\tau  = \frac{ t_\text{QSL}  }{2}$. For the zero detune $\Delta \omega=0$, the terminal points of two trajectories with $\delta \omega = 0$ and 0.1 [blue curves in Fig.~\ref{fig:detune_protocol_RWA_picture}(a)] are both very close to the equator of Bloch sphere. Their projections on $|0\rangle$ are close [blue curves of Fig.~\ref{fig:detune_protocol_RWA_picture}(b)], indicating a low sensitivity for the field strength of  $\delta \omega = 0.1$.
When using the optimal detune where $\Delta \omega \sim 0.326$, the terminal point of $\delta \omega = 0.1$ trajectory is close to equator [dash red curve in Fig.~\ref{fig:detune_protocol_RWA_picture}(a)] but that of $\delta \omega = 0$ trajectory is not [solid red curve in Fig.~\ref{fig:detune_protocol_RWA_picture}(a)]. Their projections on $|0\rangle$ are more distinct [red curves of Fig.~\ref{fig:detune_protocol_RWA_picture}(b)], indicating an enhanced sensitivity for the field strength of  $\delta \omega = 0.1$.

\subsection{Short interrogation time $\tau < T^*$ II: full calculation}
\label{sec:detune_opt}

\begin{figure}[ht]
\begin{center}
\includegraphics[width=0.95\textwidth]{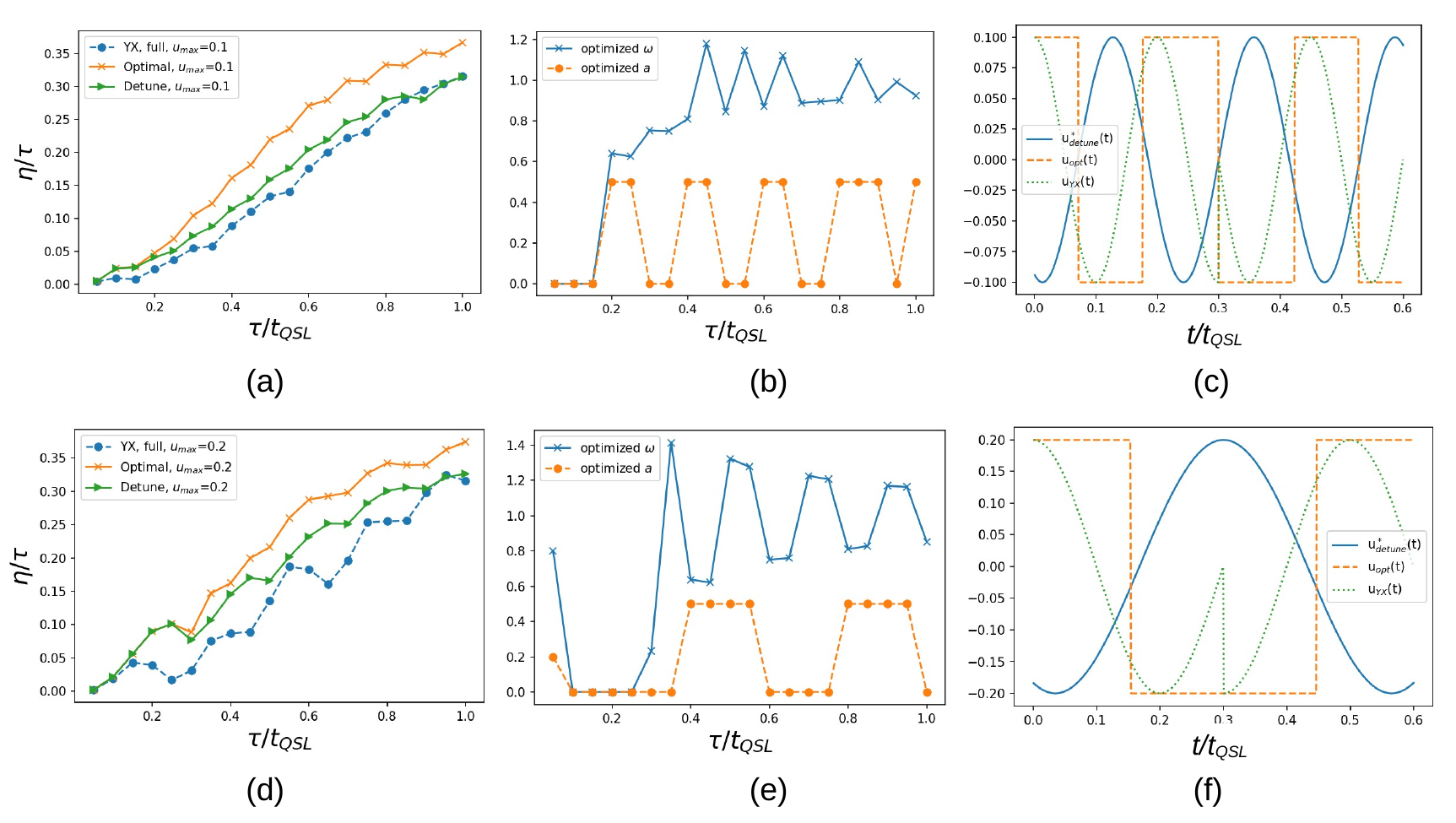}
\caption{
$\eta/\tau$ obtained from the optimal, YX, and detune protocols as a function of $\tau$ ($\tau \in [0, t_\text{QSL}]$) for $u_\text{max}=0.1$ (a) and 0.2 (d) using full calculation. The optimized detune parameters are given in (b) and (e). Notice that the $a$ is either 0 or 0.5. The corresponding optimal, YX, and detune protocol for $\tau = 0.6 \, t_\text{QSL}$ are provided in (c) and (e). The main advantage of the detune protocol is its smoothness over the entire interrogation time.
}
\label{fig:short_tau_0.6}
 \end{center}
\end{figure}

We now use full calculation to obtain the best detune and optimal protocols in the $\tau < t_\text{QSL}$ regime. With the full calculation, we re-parametrize the detune protocol  [completely equivalent to Eq.~\eqref{eqn:detune_protocol_basic}] as
\beq
\begin{aligned}
& u_\text{D}(t; \omega, a) = u_\text{max} \sin \big( \omega (t-\frac{\tau}{2}) + a \pi  \big)  \\
\Rightarrow \, &
\begin{cases}
\frac{\partial u(t)}{\partial  \omega }
= -u_\text{max} \sin \big( \omega(t-\frac{\tau}{2}) + a \pi    \big) \cdot (t-\frac{\tau}{2}) \\
\frac{\partial u(t)}{\partial  a }
= -u_\text{max} \sin \big( \omega (t-\frac{\tau}{2}) + a \pi   \big) \cdot \pi
\end{cases}.
\end{aligned}
\label{eqn:detune_parametrization}
\eeq
The applied frequency $\omega$ and a phase shift $a \pi$ are two tuning parameters. The gradients $\frac{\partial \mathcal{C}_{\eta^2} }{\partial \omega}$, $\frac{\partial \mathcal{C}_{\eta^2} }{\partial a}$ are computed using Eq.~\eqref{eqn:gradeint_alpha} and the second line of Eq.~\eqref{eqn:detune_parametrization}. The optimized $(\omega, a)$ for $u_\text{max} $ = 0.1, 0.2 are shown in Fig.~\ref{fig:short_tau_0.6}(b), (e).
We stress again that the detune protocol has the identical performance for blue and red detunes in RWA; this is not true anymore in full calculation.
Fig.~\ref{fig:short_tau_0.6}(a), (d) show the sensitivity in $\frac{\eta}{\tau}$ obtained from the optimal, YX, and detune protocols using the full calculation. Certainly the optimal protocol has the largest $|\eta|$; the optimized detune protocol performs slightly better than the reference YX protocol. In terms of control waveform, the optimal protocol is BB which has a few discontinuities; the YX protocol has one discontinuity at the midpoint $t = \frac{\tau}{2}$; the detune protocol is smooth during the entire interrogation [see Fig.~\ref{fig:short_tau_0.6}(c), (f)].

\subsection{Long interrogation time $\tau > T^*$} \label{sec:long_tau}

\begin{figure}[ht]
\begin{center}
\includegraphics[width=0.7\textwidth]{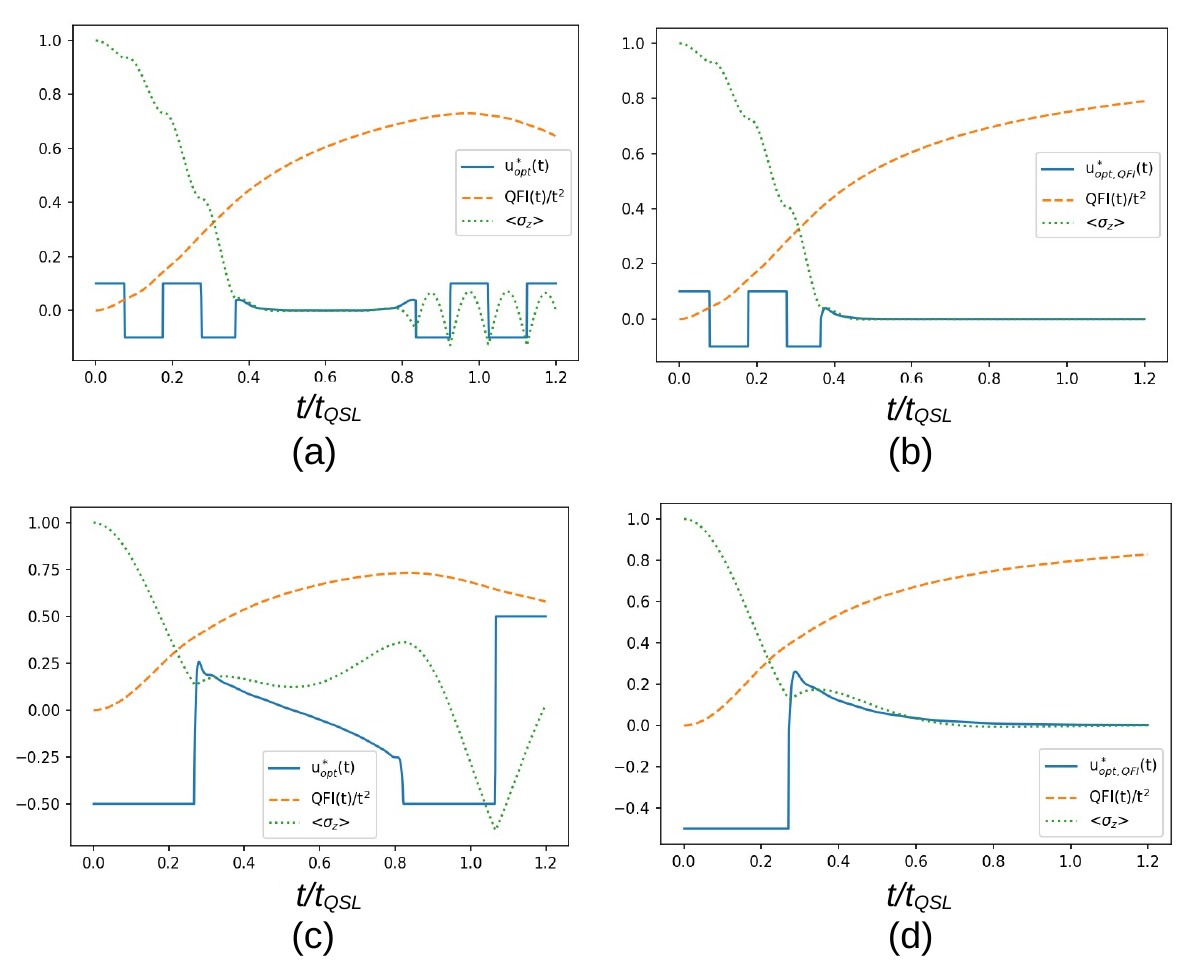}
\caption{ Optimal controls that minimize $\mathcal{C}_{\eta^2}$ and $\mathcal{C}_\text{QFI}$ for $u_\text{max} = 0.1$ (a), (b) and $u_\text{max} = 0.5$ (c), (d). The corresponding QFI$/t^2$ and $\langle \hat{\sigma}_z \rangle(t)$ are also present. The interrogation time is $\frac{\tau}{t_\text{QSL}} = 1.2$. The optimal control involves a singular arc in the middle, corresponding to the free evolution in the Ramsey protocol. According to $\langle \hat{\sigma}_z \rangle$, both $\eta$-optimal and QFI-optimal controls bring the quantum sensor to the equator of the Bloch sphere (i.e., $\langle \sigma_z \rangle=0$) as soon as possible. To maximize QFI, the sensor stays on the equator. To maximize $\eta$, the optimal control induces some oscillation around the equator in the last stage of the interrogation.
}
\label{fig:long_tau_1.2}
 \end{center}
\end{figure}

For completeness we consider the $\tau > T^*$ regime which is less relevant to high time resolution. In this regime the behavior is qualitatively similar to the celebrated Ramsey protocol \cite{PhysRev.78.695}. In Fig.~\ref{fig:long_tau_1.2} we show the results  for $u_\text{max} = 0.1$, 0.5 with the interrogation time $\tau = 1.2 \, t_\text{QSL}$.

The optimal control can be divided into three stages. In the first stage the optimal control brings the quantum sensor to the equal-superposition of two natural eigenstates (equator on the Bloch sphere) using BB waveform in a very short time; this corresponds to the first $\frac{\pi}{2}$ pulse in Ramsey protocol and is demonstrated by the reduction of $\langle \sigma_z \rangle(t)$ from 1 to (near) 0 shown in Fig.~\ref{fig:long_tau_1.2}(a), (c). The second stage is an S control which corresponds to the free evolution in the Ramsey protocol. One notes that in the optimal protocol the S control is not necessarily zero; particularly it can be significantly different from zero for large $u_\text{max}$ as shown in Fig.~\ref{fig:long_tau_1.2}(c). The final stage is again a BB control that maximizes the projection difference with and without $\delta \omega$. Overall for small $\frac{ u_\text{max} }{\omega_0}$, the optimal protocol does behave like the Ramsey protocol; the main difference is that the initial and final $\frac{\pi}{2}$ pulses in the Ramsey sequence are replaced by the faster BB protocols \cite{PhysRevA.111.042602}.

We also compute the QFI [defined in Eq.~\eqref{eqn:pi_boundary_QFI}] that quantifies the sensing performance {\em without} specifying the measurement (i.e., with best possible measurement) and has an maximum value of $t^2$ for a two-level system  \cite{PhysRevLett.96.010401, PhysRevLett.124.060402}. We find that QFI/$t^2$ does not monotonically increase but slightly decreases in the final stage. For comparison, the protocols based on optimizing QFI are given in Fig.~\ref{fig:long_tau_1.2}(b), (d) \cite{PhysRevA.103.052607}. The QFI-optimized control brings the quantum sensor to the equator of the Bloch sphere and stops, as the state at the equator is indeed most sensitive to $\delta \omega \, \hat{\sigma}_z$; the resulting QFI/$t^2$ keeps on increasing during the entire interrogation [dashed curves in Fig.~\ref{fig:long_tau_1.2}(b), (d)].  However maximizing the QFI does not take the measurement into account, and in this case, it actually leads to zero sensitivity.  This illustrates that QFI alone may not be sufficient to faithfully characterize the sensing performance under realistic conditions.

\section{Practical consideration}

\subsection{Smoothness and deviation from optimal performance} \label{sec:smoothness}

\begin{figure}[ht]
\begin{center}
\includegraphics[width=0.8\textwidth]{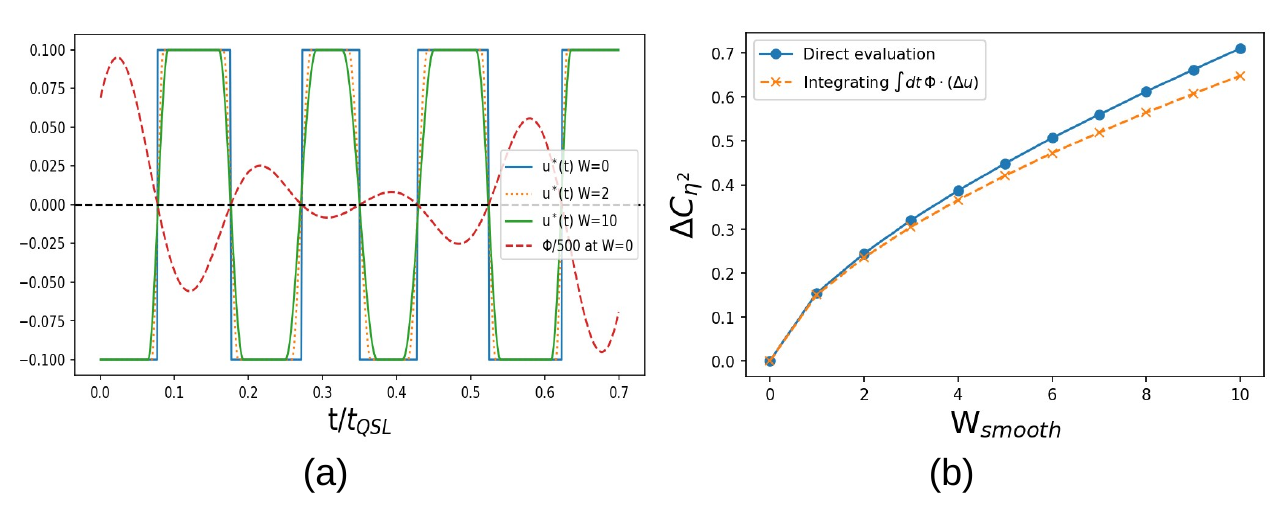}
\caption{ Imposing smoothness and estimation of performance reduction. $u_\text{max}=0.1$ is used in the calculation.
(a) Optimal controls by minimizing  Eq.~\eqref{eqn:combined_cost} using $W=0,2,10$. The switching function $\Phi$ is computed at $W_\text{smooth}=0$.
(b) The reduction of $\eta^2$ due to imposing smoothness, i.e., controls are obtained by minimizing $\mathcal{C}_\text{tot}$ with $W_\text{smooth}$ from 0 to 10.  The difference between the direct evaluation and the first order approximation [last line of Eq.~\eqref{eqn:C_reduction_estimation}] is within 10\%.
}
\label{fig:smoothness_deviation}
 \end{center}
\end{figure}

In this section, we  address two practical issues related to short-$\tau$ time-resolution protocols. The short-$\tau$ optimal protocol is purely BB which contains discontinuities at switching times, and one can smooth the protocol by introducing a running cost that imposes smoothness. 
The degree of smoothness can be quantified by \cite{PhysRevA.111.042602}
\beq
\begin{aligned}
& \mathcal{C}_\text{smooth} [ u(t) ] = \frac{1}{2} \int_0^\tau \dd t \, \dot{u}^2(t)  \\
\Rightarrow &
\frac{\delta \mathcal{C}_\text{smooth} }{ \delta u(t) } = - \dd t \, \ddot{u}.
\end{aligned}
\eeq
The Neumann boundary condition $\dot{u}(0) = \dot{u}(\tau) = 0$ is used to determine $\ddot{u} $ at two end points. To promote smoothness, the protocol is obtained by minimizing the total cost defined as
\beq
\mathcal{C}_\text{tot} = \mathcal{C}_{\eta^2} + W_\text{smooth} \mathcal{C}_\text{smooth},
\label{eqn:combined_cost}
\eeq
where the weight $W_\text{smooth}$ indicates the relative importance of smoothness. The resulting controls for $W_\text{smooth} = 0, 2, 10$ are shown in Fig.~\ref{fig:smoothness_deviation}(a). Compared to the optimal BB protocol, imposing smoothness apparently smoothens the abrupt BB waveform, but the switching times (defined by $u(t)=0$) are almost unchanged.

The performance reduction resulting from smoothness can be estimated using Eq.~\eqref{eqn:deviation_control}; in this case,
\beq
\begin{aligned}
\Delta \mathcal{C}_{\eta^2} & \equiv \mathcal{C}_{\eta^2}[ u_\text{smooth}(t) ]  - \mathcal{C}_{\eta^2}[ u^*(t) ] \\
& \approx \int \dd t \, \Phi(t) [ u_\text{smooth} (t) - u^*(t) ]
\end{aligned}
\label{eqn:C_reduction_estimation}
\eeq
We have computed the reduction by directly evaluating $\mathcal{C}_{\eta^2}$ using two controls (first line of Eq.~\eqref{eqn:C_reduction_estimation}) and by integrating over the product of switching function and the control difference [last line of Eq.~\eqref{eqn:C_reduction_estimation}]. Their difference for $W_\text{smooth} \in [0,10]$ is shown in Fig.~\ref{fig:smoothness_deviation}(b) and a very good agreement ($< 10\%$ difference) is seen. To obtain the reduction in sensitivity $\eta$, one uses $\Delta \mathcal{C}_{\eta^2} = D \big[ -\frac{ \eta^2 }{ 2} \big] = - |\eta| (\Delta \eta)$ to get $\Delta \eta = - \frac{ \Delta \mathcal{C}_{\eta^2} }{ |\eta| }$.

\subsection{Application to superconducting qubit } \label{sec:Dicussion_appl}

\begin{figure}[ht]
\begin{center}
\includegraphics[width=0.45\textwidth]{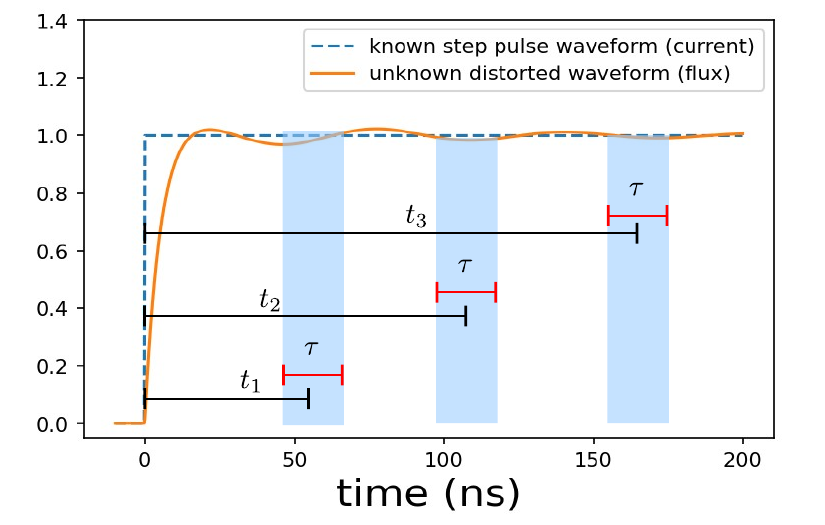}
\caption{
Potential application in calibrating the Z pulse distortion in superconducting qubits. The distortion can have time constants ranging from tens of nanoseconds to microseconds \cite{10.1063/1.5133894, hellings2025calibratingmagneticfluxcontrol, PhysRevApplied.23.024059}. The desired waveform is a step pulse (known) shown in blue and the actual waveform experienced by the qubit is a distorted pulse (unknown) shown in orange. Both are shown in arbitrary unit.
To calibrate the distortion, we first need to reconstruct the distorted waveform. We apply the short-$\tau$ time-resolution protocol centered at a series of times $t$ following the onset of the reference waveform, repeating each measurement many times to build sufficient statistics. For $u_\text{max} = 0.4$ GHz, the time resolution can be made smaller than $\tau \sim t_\text{QSL} = \frac{\pi}{ u_\text{max} } = 8$ ns. In the plot $\tau \sim 20$ ns is used. 
}
\label{fig:application}
 \end{center}
\end{figure}

As stated at the end of Section \ref{sec:overview}, employing the time-resolution protocol requires a system capable of repeatedly generating the \emph{identical}, yet unknown time-domain signal to be measured. Generating identical time-domain signals implies the ability of preparing the system of interest in exactly the {\em same} (but not necessarily known) initial state, which is the primary factor limiting its practical application. For example if the goal is to measure the magnetic domain wall propagation using the time-resolution protocol \cite{Herb_2025}, one in principle needs to repeatedly prepare the system with a magnetic domain structure with nanometer-scale precision, which can be challenging. With the hardware constraints in mind, in the following we outline a potentially impactful application.


In the context of superconducting qubits, baseband flux pulses are widely used to implement quantum gates such as the physical $Z$ gate, the iSWAP gate, and the controlled-phase (CPHASE) gate~\cite{PhysRevX.12.011005, negirneac_high_fidelity_2021, PhysRevApplied.3.034004, barends_superconducting_2014, Krantz_2019, yskp-mfcr, PhysRevApplied.18.034027, Collodo2020, chu_coupler_assisted_2021, kelly_prl_2014, PhysRevX.11.021058, PhysRevApplied.19.054050}. A long-standing challenge in these flux-based control schemes is \emph{pulse distortion} -- the deviation between the intended flux waveform and the actual waveform experienced by the qubit, arising from issues such as filtering effects and impedance mismatches in the control lines.
To address this, we propose using the time-resolution protocol (preferably the detune variant) to characterize and compensate for these distortions.
The procedure is illustrated in Fig.~\ref{fig:application}. In the example shown, the desired waveform is a known step pulse (blue, dashed curve), while the actual waveform experienced by the qubit is an \emph{unknown}, distorted pulse (orange, solid curve). Good distortion calibration necessitates an accurate reconstruction of the distorted waveform. We apply the short-$\tau$ time-resolution protocol centered at a series of times $t$ following the onset of the reference waveform, repeating each measurement many times to build sufficient statistics.
Moreover since the commonly used dispersive readout of superconducting qubits belongs to the Quantum nondemolition (QND) measurement \cite{PhysRevA.69.062320, RevModPhys.93.025005, Lupascu-2007}, i.e., the qubit remains in one of natural eigenstates following each short-$\tau$ measurement, we can sequentially probe the {\em same} distorted waveform at different values of $t$, assuming the protocol does not significantly alter the profile of the distorted pulse.
A complete calibration procedure using sequential measurements could substantially reduce the time required to collect sufficient data, thereby accelerating the overall calibration process.

To ensure clarity, the following three identifications are made for this application:
the system of interest is the room-temperature-to-cryogenic electronics; the identical, unknown time-domain signal is the flux waveform on the cryogenic side generated by well controlled room-temperature electronics; the superconducting qubit, eventually controlled by the room-temperature electronics, functions as the sensor.
Regardless of the sensor type (e.g., atoms, NV centers in diamond, ... etc), the time-resolution protocol appears suitable for calibrating known or distorted fields generated by the electronic devices, as the well-designed electronics can reliably produce identical signals. The justification for its use depends on the task at hand and the properties of the quantum sensor. For example the smallness of a quantum sensor enables detection at a precisely defined position, which is not easy or even possible for any other sensors. In this regard, using a qubit as a sensor to calibrate its own time-dependent control field is a natural fit, as it is inherently positioned in the optimal location.

\section{Conclusion}

We apply OCT to analyze the optimal performance of time resolution in quantum sensing. The performance of time resolution is quantified by the sensitivity with respect to the external field to be measured in a short interrogation time $\tau$. For this task, an end-to-end description of the sensor state over the entire measurement process is essential as any operation contributes to the interrogation time. In particular, the initial state and the basis of projection measurement have to be in the natural eigenstates of the quantum sensor. One theoretical/numerical complication arising from time resolution when applying OCT is that the terminal cost involves the derivative of the quantum state, and we cope with this additional dependence by augmenting the dynamics. This procedure not only facilitates the numerics but keeps the control system time-invariant and control-affine so that the optimality conditions from OCT remains valid. 

Similar to the reference YX protocol proposed in Ref.~\cite{PhysRevLett.133.210802}, the behavior of optimal protocols exhibits a critical interrogation time $T^*$: when $\tau <T^*$ the optimal protocol is purely bang-bang; when $\tau>T^*$ the optimal protocol involves a singular control the interrogation. The latter corresponds to the free evolution in the Ramsey protocol. Given the same amplitude constraint $u_\text{max}$, the optimal control achieves a higher sensitivity over the reference protocol. The metrological gain is more substantial in the regime of short interrogation time, the regime relevant to high time resolution. 

Despite the metrological gain in the short-$\tau$ regime, the bang-bang waveform of the optimal control contains several discontinuities, making it experimentally challenging to implement. The reference YX protocol is smoother but still carries one discontinuity at the midpoint of the interrogation. To address this issue, we propose a ``detune protocol'' that is smooth during the entire interrogation with performance comparable to (slightly better than) that of the reference YX protocol. An analytical approximation within RWA is derived, and physical intuition is developed based on Bloch sphere trajectories. We expect the detune protocol to be practically useful for the short-$\tau$ time-resolution protocol. In the long-$\tau$ regime, we evaluate QFI and construct the QFI-maximized protocols to illustrate the difference between theoretically optimal measurements (as indicated by QFI) and those that are practically realizable.

To meaningfully utilize the time-resolution protocol requires a setting in which the \emph{identical}, yet unknown, time-domain signal can be generated repeatedly.
As a potentially promising application, we propose using the time-resolution protocol to calibrate the waveform distortion between the intended flux waveform and the actual waveform experienced by a superconducting qubit. In this context, the superconducting qubit functions as a sensor. 
The advantages specific to this application, notably the QND-enabled sequential measurements, are discussed. We hope and expect that the proposed protocol will be useful for calibrating a qubit's control field.

\section*{Acknowledgment} 
We thank Junyoung An (MIT) for very insightful comments. C.L. thanks helpful discussions from Dries Sels (New York University and Flatiron Institute) and Yen-Hsiang Lin (National Tsing Hua University, Taiwan), and  Chih-Chun Chien (University of California at Merced).
Q. D. is grateful to William D. Oliver and Jeffery A. Grover for their support and helpful discussions.
\appendix

\section{Kernel function construction} \label{app:kernel}

\begin{figure}[ht]
\begin{center}
\includegraphics[width=0.45\textwidth]{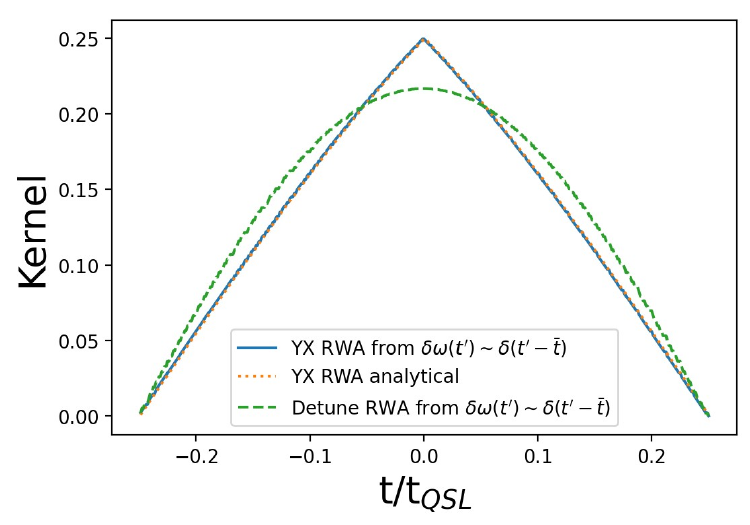}
\caption{ RWA evaluated kernel functions [Eq.~\eqref{eqn:convolution}] for YX and detune protocols. $u_\text{max}=0.5$ and  $\tau = 0.5 t_\text{QSL}$ are used; the detune value is $\sim 0.813$ using Eq.~\eqref{eqn:optimal_detune_RWA_approx}. Numerical kernel function is constructed by applying a series of  $\delta \omega(t') = \delta(t' -\bar{t})$ with 200 $\bar{t}$'s. For the YX protocol it matches well with the analytical expression [Eq.~\eqref{eqn:Kernel_YX_RWA_analytical}]. The kernel for the detune protocol is smoother. The integrated areas of both protocols are comparable because they have similar sensitivity values.
}
\label{fig:Kernel}
 \end{center}
\end{figure}

In this Appendix we provide a straightforward way to construct the kernel $K$ in Eq.~\eqref{eqn:convolution}. 
Without loss of generality we take $t=0$ to simplify the notation. $K(t')$ vanishes when $|t'|> \frac{\tau}{2}$ because during these periods the quanutm sensor is in one of natural eigenstates (i.e., eigenstates of $\hat{\sigma}_z$) that does not respond to $\delta \omega \, \hat{\sigma}_z$. When $\delta \omega(t') = \delta \omega$ is a constant over $|t'| < \frac{\tau}{2}$, the integrated area $\int \dd t' K(t') = \eta$ .

The analytical expression for kernel function is not easy to derive if exists, but can be numerically constructed. The most straightforward way is to apply a sequence of  $\delta \omega(t') = \delta (t' - \bar{t})$ for $\bar{t} \in [-\frac{\tau}{2},  +\frac{\tau}{2}]$, and at each $\bar{t}$ the resulting $\delta p(0) = K(\bar{t})$.  We test this method by constructing $K$ of the YX protocol using RWA, and the profile agrees very well with the analytical expression
\beq
K(t') = \frac{\Theta(\frac{\tau}{2}-|t'|)}{2} \sin (\frac{u_\text{max} \tau}{2} ) \cdot \sin \big( u_\text{max} ( \frac{\tau}{2} - |t'| ) \big)
\label{eqn:Kernel_YX_RWA_analytical}
\eeq
given in Ref.~\cite{PhysRevLett.133.210802}.
Results for $\tau = 0.5 \, t_\text{QSL}$ for YX and detune protocols using RWA are given in Fig.~\ref{fig:Kernel}. It is seen that the kernel of the detune protocol is smoother. The integrated areas of both protocols are comparable, indicating they have similar sensitivity values. 

\section{Second-order condition for singular control} \label{App:singular_value}

Singular control $u^\text{sing}(t)$ given in Eq.~\eqref{eqn:singular_control_value} is determined by requiring $\Phi(t) = \dot{\Phi} = \ddot{\Phi} = 0$. In the following equations we only take the real part on the right-hand side:
\beq
\begin{aligned}
 \dot{\Phi} &=  (-i)^2  \langle \pi | \big[ \begin{bmatrix} \sigma_x &0 \\ 0 & \sigma_x \end{bmatrix} , \begin{bmatrix} \frac{\omega_0}{2} \sigma_z &0 \\ \frac{1}{2} \sigma_z & \frac{\omega_0}{2} \sigma_z \end{bmatrix}     \big] | \psi \rangle
  = +  2i \langle \pi | \begin{bmatrix} \frac{\omega_0}{2} \sigma_y &0 \\ \frac{1}{2} \sigma_y & \frac{\omega_0}{2} \sigma_y \end{bmatrix} | \psi \rangle   \\
  \ddot{\Phi} 
  &= 4 i \bigg( \frac{\omega_0^2}{4} \Phi(t) + \frac{\omega_0}{2}  \langle \pi | \begin{bmatrix} 0&0 \\ \sigma_x &0 \end{bmatrix}  | \psi \rangle -
  u^\text{sing}(t) (t) \langle \pi | \begin{bmatrix} \frac{\omega_0}{2} \sigma_z &0 \\ \frac{1}{2} \sigma_z & \frac{\omega_0}{2} \sigma_z \end{bmatrix}   | \psi \rangle   \bigg) = 0.
\end{aligned}
\label{eqn:singular_control_2nd_order}
\eeq
Using Re[$-i A$] = Im[$A$] leads to Eq.~\eqref{eqn:singular_control_value}.

\bibliography{TimeResolution}

@book{book:Luenberger,
   title =     {Introduction to dynamic systems: theory, models, and applications},
   author =    {David G. Luenberger},
   publisher = {Wiley, New York},
   year =      {1979}
}

@book{book:Liberzon,
   title =     {Calculus of Variations and Optimal Control Theory: A Concise Introduction},
   author =    {Daniel Liberzon},
   publisher = {Princeton University Press},
   year =      {2012}
}

@book{book:Helstrom,
   title =     {Quantum Detection and Estimation Theory},
   author =    {Carl W. Helstrom },
   publisher = {Elsevier, Academic Press},
   year =      {1976},
   series =    {Mathematics in Science and Engineering 123}
}

@book{book:Holevo,
   title =     {Probabilistic and Statistical Aspects of Quantum Theory},
   author =    {Alexander S. Holevo},
   publisher = {Edizioni della Normale},
   year =      {2011},
   edition =   {1st Edition}
}

@misc{QuantumSpeedLimit, 
note = { The quantum speed limit in this context refers to the minimum time required to guide the qubit from $|0\rangle$ or $|1 \rangle$  (i.e., one of poles on the Bloch sphere) to the state of equal superposition of $|0\rangle$ and $|1 \rangle$ (i.e., the equator on the Bloch sphere) given $u(t) = u_\text{max} \cos(\omega_0 t)$ using RWA.   }
}

@misc{control_type, 
note = {Rigorously these two optimality conditions hold only when the control dynamics is control-affine [linear in $u(t)$] and time-invariant [the time dependence is solely from $u(t)$]. The augmented dynamics of Eq.~\eqref{eqn:forward} meets these two conditions.   }
}

@misc{smith2025interradicalmotionpushmagnetosensing,
      title={Interradical motion can push magnetosensing precision towards quantum limits}, 
      author={Luke D. Smith and Farhan T. Chowdhury and Jonas Glatthard and Daniel R. Kattnig},
      year={2025},
      eprint={2506.21389},
      archivePrefix={arXiv},
      primaryClass={quant-ph},
      url={https://arxiv.org/abs/2506.21389}, 
}

@article{doi:10.1021/acs.jpclett.2c02840,
author = {Smith, Luke D. and Chowdhury, Farhan T. and Peasgood, Iona and Dawkins, Nahnsu and Kattnig, Daniel R.},
title = {Driven Radical Motion Enhances Cryptochrome Magnetoreception: Toward Live Quantum Sensing},
journal = {The Journal of Physical Chemistry Letters},
volume = {13},
number = {45},
pages = {10500-10506},
year = {2022},
doi = {10.1021/acs.jpclett.2c02840},
    note ={PMID: 36332112}, 
URL = {  https://doi.org/10.1021/acs.jpclett.2c02840 },
eprint = { https://doi.org/10.1021/acs.jpclett.2c02840}
}

@article{PRXQuantum.5.020303,
  title = {Quantum Control of Radical-Pair Dynamics beyond Time-Local Optimization},
  author = {Chowdhury, Farhan T. and Denton, Matt C.J. and Bonser, Daniel C. and Kattnig, Daniel R.},
  journal = {PRX Quantum},
  volume = {5},
  issue = {2},
  pages = {020303},
  numpages = {18},
  year = {2024},
  month = {Apr},
  publisher = {American Physical Society},
  doi = {10.1103/PRXQuantum.5.020303},
  url = {https://link.aps.org/doi/10.1103/PRXQuantum.5.020303}
}

@article{PhysRevLett.104.220502,
  title = {Quantum Control and Entanglement in a Chemical Compass},
  author = {Cai, Jianming and Guerreschi, Gian Giacomo and Briegel, Hans J.},
  journal = {Phys. Rev. Lett.},
  volume = {104},
  issue = {22},
  pages = {220502},
  numpages = {4},
  year = {2010},
  month = {Jun},
  publisher = {American Physical Society},
  doi = {10.1103/PhysRevLett.104.220502},
  url = {https://link.aps.org/doi/10.1103/PhysRevLett.104.220502}
}

@article{RevModPhys.89.035002,
  title = {Quantum sensing},
  author = {Degen, C. L. and Reinhard, F. and Cappellaro, P.},
  journal = {Rev. Mod. Phys.},
  volume = {89},
  issue = {3},
  pages = {035002},
  numpages = {39},
  year = {2017},
  month = {Jul},
  publisher = {American Physical Society},
  doi = {10.1103/RevModPhys.89.035002},
  url = {https://link.aps.org/doi/10.1103/RevModPhys.89.035002}
}

@article{Herb_2025,
   title={Quantum magnetometry of transient signals with a time resolution of 1.1 nanoseconds},
   volume={16},
   ISSN={2041-1723},
   url={http://dx.doi.org/10.1038/s41467-025-55956-1},
   DOI={10.1038/s41467-025-55956-1},
   number={1},
   journal={Nature Communications},
   publisher={Springer Science and Business Media LLC},
   author={Herb, K. and V\"olker, L. A. and Abendroth, J. M. and Meinhardt, N. and van Schie, L. and Gambardella, P. and Degen, C. L.},
   year={2025},
   month=jan }

@article{10.1063/1.5133894,
    author = {Rol, M. A. and Ciorciaro, L. and Malinowski, F. K. and Tarasinski, B. M. and Sagastizabal, R. E. and Bultink, C. C. and Salathe, Y. and Haandbaek, N. and Sedivy, J. and DiCarlo, L.},
    title = {Time-domain characterization and correction of on-chip distortion of control pulses in a quantum processor},
    journal = {Applied Physics Letters},
    volume = {116},
    number = {5},
    pages = {054001},
    year = {2020},
    month = {02},
    issn = {0003-6951},
    doi = {10.1063/1.5133894},
    url = {https://doi.org/10.1063/1.5133894},
    eprint = {https://pubs.aip.org/aip/apl/article-pdf/doi/10.1063/1.5133894/14531145/054001\_1\_online.pdf},
}

@book{book:GeometricOptimalControl,
   title =     {Geometric Optimal Control: Theory, Methods and Examples},
   author =    {Heinz Schattler, Urszula Ledzewicz},
   publisher = {Springer-Verlag New York},
   isbn =      {978-1-4614-3833-5,978-1-4614-3834-2},
   year =      {2012},
   series =    {Interdisciplinary Applied Mathematics 38},
   edition =   {1}
}

@book{book:Pontryagin,
 author = {L. S. Pontryagin}, 
  title = {Mathematical Theory of Optimal Processes},
 publisher = {CRC Press, Boca Raton, FL},
  year = {1987}
}

@article{PhysRevLett.133.210802,
  title = {Quantum Speed Limit in Quantum Sensing},
  author = {Herb, K. and Degen, C. L.},
  journal = {Phys. Rev. Lett.},
  volume = {133},
  issue = {21},
  pages = {210802},
  numpages = {6},
  year = {2024},
  month = {Nov},
  publisher = {American Physical Society},
  doi = {10.1103/PhysRevLett.133.210802},
  url = {https://link.aps.org/doi/10.1103/PhysRevLett.133.210802}
}

@article{Mandelstam_1945,
title	= {The uncertainty relation between energy and time in nonrelativistic quantum mechanics},
author	= {Mandelstam, L and Tamm, I.},
journal	= {J. Phys. (USSR) },
year	= {1945},
volume	= {9},
pages	= {249}
}

@article{Norma_1998,
doi	= {10.1016/s0167-2789(98)00054-2},
title	= {The maximum speed of dynamical evolution},
author	= {Norman Margolus and Lev B. Levitin},
journal	= {Physica D},
year	= {1998},
month	= {sep},
volume	= {120},
issue	= {1-2},
pages	= {188}
}

@article{Deffner_2017,
doi = {10.1088/1751-8121/aa86c6},
url = {https://dx.doi.org/10.1088/1751-8121/aa86c6},
year = {2017},
month = {oct},
publisher = {IOP Publishing},
volume = {50},
number = {45},
pages = {453001},
author = {Deffner, Sebastian and Campbell, Steve},
title = {Quantum speed limits: from Heisenberg’s uncertainty principle to optimal quantum control},
journal = {Journal of Physics A: Mathematical and Theoretical}
}

@article{PhysRevLett.124.060402,
  title = {Machine-Designed Sensor to Make Optimal Use of Entanglement-Generating Dynamics for Quantum Sensing},
  author = {Haine, Simon A. and Hope, Joseph J.},
  journal = {Phys. Rev. Lett.},
  volume = {124},
  issue = {6},
  pages = {060402},
  numpages = {6},
  year = {2020},
  month = {Feb},
  publisher = {American Physical Society},
  doi = {10.1103/PhysRevLett.124.060402},
  url = {https://link.aps.org/doi/10.1103/PhysRevLett.124.060402}
}

@article{PhysRevLett.59.278,
  title = {Precision measurement beyond the shot-noise limit},
  author = {Xiao, Min and Wu, Ling-An and Kimble, H. J.},
  journal = {Phys. Rev. Lett.},
  volume = {59},
  issue = {3},
  pages = {278--281},
  numpages = {0},
  year = {1987},
  month = {Jul},
  publisher = {American Physical Society},
  doi = {10.1103/PhysRevLett.59.278},
  url = {https://link.aps.org/doi/10.1103/PhysRevLett.59.278}
}

@article{PhysRevLett.59.2153,
  title = {Squeezed-light--enhanced polarization interferometer},
  author = {Grangier, P. and Slusher, R. E. and Yurke, B. and LaPorta, A.},
  journal = {Phys. Rev. Lett.},
  volume = {59},
  issue = {19},
  pages = {2153--2156},
  numpages = {0},
  year = {1987},
  month = {Nov},
  publisher = {American Physical Society},
  doi = {10.1103/PhysRevLett.59.2153},
  url = {https://link.aps.org/doi/10.1103/PhysRevLett.59.2153}
}

@article{Breitenbach-1997,
doi	= {10.1038/387471a0},
title	= {Measurement of the quantum states of squeezed light},
author	= {Breitenbach, G. and Schiller, S. and  Mlynek, J.},
journal	= {Nature},
year	= {1997},
month	= {may},
volume	= {387},
issue	= {6632},
page	= {471--475}
}

@misc{hellings2025calibratingmagneticfluxcontrol,
      title={Calibrating Magnetic Flux Control in Superconducting Circuits by Compensating Distortions on Time Scales from Nanoseconds up to Tens of Microseconds}, 
      author={Christoph Hellings and Nathan Lacroix and Ants Remm and Richard Boell and Johannes Herrmann and Stefania Lazăr and Sebastian Krinner and François Swiadek and Christian Kraglund Andersen and Christopher Eichler and Andreas Wallraff},
      year={2025},
      eprint={2503.04610},
      archivePrefix={arXiv},
      primaryClass={quant-ph},
      url={https://arxiv.org/abs/2503.04610}, 
}

@article{PhysRevApplied.23.024059,
  title = {High-precision pulse calibration of tunable couplers for high-fidelity two-qubit gates in superconducting quantum processors},
  author = {Li, Tian-Ming and Zhang, Jia-Chi and Chen, Bing-Jie and Huang, Kaixuan and Liu, Hao-Tian and Xiao, Yong-Xi and Deng, Cheng-Lin and Liang, Gui-Han and Chen, Chi-Tong and Liu, Yu and Li, Hao and Bao, Zhen-Ting and Zhao, Kui and Xu, Yueshan and Li, Li and He, Yang and Liu, Zheng-He and Yu, Yi-Han and Zhou, Si-Yun and Liu, Yan-Jun and Song, Xiaohui and Zheng, Dongning and Xiang, Zhongcheng and Shi, Yun-Hao and Xu, Kai and Fan, Heng},
  journal = {Phys. Rev. Appl.},
  volume = {23},
  issue = {2},
  pages = {024059},
  numpages = {16},
  year = {2025},
  month = {Feb},
  publisher = {American Physical Society},
  doi = {10.1103/PhysRevApplied.23.024059},
  url = {https://link.aps.org/doi/10.1103/PhysRevApplied.23.024059}
}

@article{Li_2025,
   title={High-precision pulse calibration of tunable couplers for high-fidelity two-qubit gates in superconducting quantum processors},
   volume={23},
   ISSN={2331-7019},
   url={http://dx.doi.org/10.1103/PhysRevApplied.23.024059},
   DOI={10.1103/physrevapplied.23.024059},
   number={2},
   journal={Physical Review Applied},
   publisher={American Physical Society (APS)},
   author={Li, Tian-Ming and Zhang, Jia-Chi and Chen, Bing-Jie and Huang, Kaixuan and Liu, Hao-Tian and Xiao, Yong-Xi and Deng, Cheng-Lin and Liang, Gui-Han and Chen, Chi-Tong and Liu, Yu and Li, Hao and Bao, Zhen-Ting and Zhao, Kui and Xu, Yueshan and Li, Li and He, Yang and Liu, Zheng-He and Yu, Yi-Han and Zhou, Si-Yun and Liu, Yan-Jun and Song, Xiaohui and Zheng, Dongning and Xiang, Zhongcheng and Shi, Yun-Hao and Xu, Kai and Fan, Heng},
   year={2025},
   month=feb }

@article{PhysRevX.11.021058,
  title = {Realization of High-Fidelity CZ and $ZZ$-Free iSWAP Gates with a Tunable Coupler},
  author = {Sung, Youngkyu and Ding, Leon and Braum\"uller, Jochen and Veps\"al\"ainen, Antti and Kannan, Bharath and Kjaergaard, Morten and Greene, Ami and Samach, Gabriel O. and McNally, Chris and Kim, David and Melville, Alexander and Niedzielski, Bethany M. and Schwartz, Mollie E. and Yoder, Jonilyn L. and Orlando, Terry P. and Gustavsson, Simon and Oliver, William D.},
  journal = {Phys. Rev. X},
  volume = {11},
  issue = {2},
  pages = {021058},
  numpages = {32},
  year = {2021},
  month = {Jun},
  publisher = {American Physical Society},
  doi = {10.1103/PhysRevX.11.021058},
  url = {https://link.aps.org/doi/10.1103/PhysRevX.11.021058}
}

@article{Krantz_2019,
   title={A quantum engineer’s guide to superconducting qubits},
   volume={6},
   ISSN={1931-9401},
   url={http://dx.doi.org/10.1063/1.5089550},
   DOI={10.1063/1.5089550},
   number={2},
   journal={Applied Physics Reviews},
   publisher={AIP Publishing},
   author={Krantz, P. and Kjaergaard, M. and Yan, F. and Orlando, T. P. and Gustavsson, S. and Oliver, W. D.},
   year={2019}
   }

@article{barends_superconducting_2014,
	title = {Superconducting quantum circuits at the surface code threshold for fault tolerance},
	volume = {508},
	copyright = {2014 Nature Publishing Group, a division of Macmillan Publishers Limited. All Rights Reserved.},
	issn = {1476-4687},
	url = {https://www.nature.com/articles/nature13171},
	doi = {10.1038/nature13171},
	number = {7497},
	urldate = {2023-02-26},
	journal = {Nature},
	author = {Barends, R. and Kelly, J. and Megrant, A. and Veitia, A. and Sank, D. and Jeffrey, E. and White, T. C. and Mutus, J. and Fowler, A. G. and Campbell, B. and Chen, Y. and Chen, Z. and Chiaro, B. and Dunsworth, A. and Neill, C. and O’Malley, P. and Roushan, P. and Vainsencher, A. and Wenner, J. and Korotkov, A. N. and Cleland, A. N. and Martinis, John M.},
	month = apr,
	year = {2014},
	pages = {500--503},
}

@article{PhysRevApplied.3.034004,
  title = {$Z$-Gate Operation on a Superconducting Flux Qubit via its Readout SQUID},
  author = {Jin, X. Y. and Gustavsson, S. and Bylander, J. and Yan, F. and Yoshihara, F. and Nakamura, Y. and Orlando, T. P. and Oliver, W. D.},
  journal = {Phys. Rev. Appl.},
  volume = {3},
  issue = {3},
  pages = {034004},
  numpages = {6},
  year = {2015},
  month = {Mar},
  publisher = {American Physical Society},
  doi = {10.1103/PhysRevApplied.3.034004},
  url = {https://link.aps.org/doi/10.1103/PhysRevApplied.3.034004}
}

@article{PhysRevApplied.18.034027,
  title = {Fast Flux Entangling Gate for Fluxonium Circuits},
  author = {Chen, Yinqi and Nesterov, Konstantin N. and Manucharyan, Vladimir E. and Vavilov, Maxim G.},
  journal = {Phys. Rev. Appl.},
  volume = {18},
  issue = {3},
  pages = {034027},
  numpages = {12},
  year = {2022},
  month = {Sep},
  publisher = {American Physical Society},
  doi = {10.1103/PhysRevApplied.18.034027},
  url = {https://link.aps.org/doi/10.1103/PhysRevApplied.18.034027}
}

@article{PhysRevX.12.011005,
  title = {Demonstration of Density Matrix Exponentiation Using a Superconducting Quantum Processor},
  author = {Kjaergaard, M. and Schwartz, M. E. and Greene, A. and Samach, G. O. and Bengtsson, A. and O'Keeffe, M. and McNally, C. M. and Braum\"uller, J. and Kim, D. K. and Krantz, P. and Marvian, M. and Melville, A. and Niedzielski, B. M. and Sung, Y. and Winik, R. and Yoder, J. and Rosenberg, D. and Obenland, K. and Lloyd, S. and Orlando, T. P. and Marvian, I. and Gustavsson, S. and Oliver, W. D.},
  journal = {Phys. Rev. X},
  volume = {12},
  issue = {1},
  pages = {011005},
  numpages = {21},
  year = {2022},
  month = {Jan},
  publisher = {American Physical Society},
  doi = {10.1103/PhysRevX.12.011005},
  url = {https://link.aps.org/doi/10.1103/PhysRevX.12.011005}
}

@article{negirneac_high_fidelity_2021,
  title = {High-Fidelity Controlled-$Z$ Gate with Maximal Intermediate Leakage Operating at the Speed Limit in a Superconducting Quantum Processor},
  author = {Neg\^{\i}rneac, V. and Ali, H. and Muthusubramanian, N. and Battistel, F. and Sagastizabal, R. and Moreira, M. S. and Marques, J. F. and Vlothuizen, W. J. and Beekman, M. and Zachariadis, C. and Haider, N. and Bruno, A. and DiCarlo, L.},
  journal = {Phys. Rev. Lett.},
  volume = {126},
  issue = {22},
  pages = {220502},
  numpages = {6},
  year = {2021},
  month = {Jun.},
  publisher = {American Physical Society},
  doi = {10.1103/PhysRevLett.126.220502},
  url = {https://link.aps.org/doi/10.1103/PhysRevLett.126.220502}
}

@article{yskp-mfcr,
  title = {Pulse design of baseband flux control for adiabatic controlled-phase gates in superconducting circuits},
  author = {Ding, Qi and Oppenheim, Alan V. and Boufounos, Petros T. and Gustavsson, Simon and Grover, Jeffrey A. and Baran, Thomas A. and Oliver, William D.},
  journal = {Phys. Rev. Appl.},
  volume = {23},
  issue = {6},
  pages = {064013},
  numpages = {27},
  year = {2025},
  month = {Jun},
  publisher = {American Physical Society},
  doi = {10.1103/yskp-mfcr},
  url = {https://link.aps.org/doi/10.1103/yskp-mfcr}
}

@article{Collodo2020,
  title = {Implementation of Conditional Phase Gates Based on Tunable $ZZ$ Interactions},
  author = {Collodo, Michele C. and Herrmann, Johannes and Lacroix, Nathan and Andersen, Christian Kraglund and Remm, Ants and Lazar, Stefania and Besse, Jean-Claude and Walter, Theo and Wallraff, Andreas and Eichler, Christopher},
  journal = {Phys. Rev. Lett.},
  volume = {125},
  issue = {24},
  pages = {240502},
  numpages = {5},
  year = {2020},
  month = {Dec},
  publisher = {American Physical Society},
  doi = {10.1103/PhysRevLett.125.240502},
  url = {https://link.aps.org/doi/10.1103/PhysRevLett.125.240502}
}

@article{chu_coupler_assisted_2021,
	title = {Coupler-{Assisted} {Controlled}-{Phase} {Gate} with {Enhanced} {Adiabaticity}},
	volume = {16},
	url = {https://link.aps.org/doi/10.1103/PhysRevApplied.16.054020},
	doi = {10.1103/PhysRevApplied.16.054020},
	number = {5},
	urldate = {2023-02-26},
	journal = {Physical Review Applied},
	author = {Chu, Ji and Yan, Fei},
	month = nov,
	year = {2021},
	pages = {054020} 
}

@article{PhysRevApplied.19.054050,
  title = {Baseband Control of Superconducting Qubits with Shared Microwave Drives},
  author = {Zhao, Peng and Wang, Ruixia and Hu, Meng-Jun and Ma, Teng and Xu, Peng and Jin, Yirong and Yu, Haifeng},
  journal = {Phys. Rev. Appl.},
  volume = {19},
  issue = {5},
  pages = {054050},
  numpages = {18},
  year = {2023},
  month = {May},
  publisher = {American Physical Society},
  doi = {10.1103/PhysRevApplied.19.054050},
  url = {https://link.aps.org/doi/10.1103/PhysRevApplied.19.054050}
}

@article{kelly_prl_2014,
  title = {Optimal Quantum Control Using Randomized Benchmarking},
  author = {Kelly, J. and Barends, R. and Campbell, B. and Chen, Y. and Chen, Z. and Chiaro, B. and Dunsworth, A. and Fowler, A. G. and Hoi, I.-C. and Jeffrey, E. and Megrant, A. and Mutus, J. and Neill, C. and O'Malley, P. J. J. and Quintana, C. and Roushan, P. and Sank, D. and Vainsencher, A. and Wenner, J. and White, T. C. and Cleland, A. N. and Martinis, John M.},
  journal = {Phys. Rev. Lett.},
  volume = {112},
  issue = {24},
  pages = {240504},
  numpages = {5},
  year = {2014},
  month = {Jun.},
  publisher = {American Physical Society},
  doi = {10.1103/PhysRevLett.112.240504},
  url = {https://link.aps.org/doi/10.1103/PhysRevLett.112.240504}
}

@article{PhysRevLett.123.231107,
  title = {Quantum-Enhanced Advanced LIGO Detectors in the Era of Gravitational-Wave Astronomy},
  author = {Tse, M. and Yu, Haocun and Kijbunchoo, N. and Fernandez-Galiana, A. and Dupej, P. and Barsotti, L. and Blair, C. D. and Brown, D. D. and Dwyer, S. E. and Effler, A. and Evans, M. and Fritschel, P. and Frolov, V. V. and Green, A. C. and Mansell, G. L. and Matichard, F. and Mavalvala, N. and McClelland, D. E. and McCuller, L. and McRae, T. and Miller, J. and Mullavey, A. and Oelker, E. and Phinney, I. Y. and Sigg, D. and Slagmolen, B. J. J. and Vo, T. and Ward, R. L. and Whittle, C. and Abbott, R. and Adams, C. and Adhikari, R. X. and Ananyeva, A. and Appert, S. and Arai, K. and Areeda, J. S. and Asali, Y. and Aston, S. M. and Austin, C. and Baer, A. M. and Ball, M. and Ballmer, S. W. and Banagiri, S. and Barker, D. and Bartlett, J. and Berger, B. K. and Betzwieser, J. and Bhattacharjee, D. and Billingsley, G. and Biscans, S. and Blair, R. M. and Bode, N. and Booker, P. and Bork, R. and Bramley, A. and Brooks, A. F. and Buikema, A. and Cahillane, C. and Cannon, K. C. and Chen, X. and Ciobanu, A. A. and Clara, F. and Cooper, S. J. and Corley, K. R. and Countryman, S. T. and Covas, P. B. and Coyne, D. C. and Datrier, L. E. H. and Davis, D. and Di Fronzo, C. and Driggers, J. C. and Etzel, T. and Evans, T. M. and Feicht, J. and Fulda, P. and Fyffe, M. and Giaime, J. A. and Giardina, K. D. and Godwin, P. and Goetz, E. and Gras, S. and Gray, C. and Gray, R. and Gupta, Anchal and Gustafson, E. K. and Gustafson, R. and Hanks, J. and Hanson, J. and Hardwick, T. and Hasskew, R. K. and Heintze, M. C. and Helmling-Cornell, A. F. and Holland, N. A. and Jones, J. D. and Kandhasamy, S. and Karki, S. and Kasprzack, M. and Kawabe, K. and King, P. J. and Kissel, J. S. and Kumar, Rahul and Landry, M. and Lane, B. B. and Lantz, B. and Laxen, M. and Lecoeuche, Y. K. and Leviton, J. and Liu, J. and Lormand, M. and Lundgren, A. P. and Macas, R. and MacInnis, M. and Macleod, D. M. and M\'arka, S. and M\'arka, Z. and Martynov, D. V. and Mason, K. and Massinger, T. J. and McCarthy, R. and McCormick, S. and McIver, J. and Mendell, G. and Merfeld, K. and Merilh, E. L. and Meylahn, F. and Mistry, T. and Mittleman, R. and Moreno, G. and Mow-Lowry, C. M. and Mozzon, S. and Nelson, T. J. N. and Nguyen, P. and Nuttall, L. K. and Oberling, J. and Oram, R. J. and O'Reilly, B. and Osthelder, C. and Ottaway, D. J. and Overmier, H. and Palamos, J. R. and Parker, W. and Payne, E. and Pele, A. and Perez, C. J. and Pirello, M. and Radkins, H. and Ramirez, K. E. and Richardson, J. W. and Riles, K. and Robertson, N. A. and Rollins, J. G. and Romel, C. L. and Romie, J. H. and Ross, M. P. and Ryan, K. and Sadecki, T. and Sanchez, E. J. and Sanchez, L. E. and Saravanan, T. R. and Savage, R. L. and Schaetzl, D. and Schnabel, R. and Schofield, R. M. S. and Schwartz, E. and Sellers, D. and Shaffer, T. J. and Smith, J. R. and Soni, S. and Sorazu, B. and Spencer, A. P. and Strain, K. A. and Sun, L. and Szczepa\ifmmode \acute{n}\else \'{n}\fi{}czyk, M. J. and Thomas, M. and Thomas, P. and Thorne, K. A. and Toland, K. and Torrie, C. I. and Traylor, G. and Urban, A. L. and Vajente, G. and Valdes, G. and Vander-Hyde, D. C. and Veitch, P. J. and Venkateswara, K. and Venugopalan, G. and Viets, A. D. and Vorvick, C. and Wade, M. and Warner, J. and Weaver, B. and Weiss, R. and Willke, B. and Wipf, C. C. and Xiao, L. and Yamamoto, H. and Yap, M. J. and Yu, Hang and Zhang, L. and Zucker, M. E. and Zweizig, J.},
  journal = {Phys. Rev. Lett.},
  volume = {123},
  issue = {23},
  pages = {231107},
  numpages = {8},
  year = {2019},
  month = {Dec},
  publisher = {American Physical Society},
  doi = {10.1103/PhysRevLett.123.231107},
  url = {https://link.aps.org/doi/10.1103/PhysRevLett.123.231107}
}

@article{PhysRevLett.123.231108,
  title = {Increasing the Astrophysical Reach of the Advanced Virgo Detector via the Application of Squeezed Vacuum States of Light},
  author = {Acernese, F. and Agathos, M. and Aiello, L. and Allocca, A. and Amato, A. and Ansoldi, S. and Antier, S. and Ar\`ene, M. and Arnaud, N. and Ascenzi, S. and Astone, P. and Aubin, F. and Babak, S. and Bacon, P. and Badaracco, F. and Bader, M. K. M. and Baird, J. and Baldaccini, F. and Ballardin, G. and Baltus, G. and Barbieri, C. and Barneo, P. and Barone, F. and Barsuglia, M. and Barta, D. and Basti, A. and Bawaj, M. and Bazzan, M. and Bejger, M. and Belahcene, I. and Bernuzzi, S. and Bersanetti, D. and Bertolini, A. and Bischi, M. and Bitossi, M. and Bizouard, M. A. and Bobba, F. and Boer, M. and Bogaert, G. and Bondu, F. and Bonnand, R. and Boom, B. A. and Boschi, V. and Bouffanais, Y. and Bozzi, A. and Bradaschia, C. and Branchesi, M. and Breschi, M. and Briant, T. and Brighenti, F. and Brillet, A. and Brooks, J. and Bruno, G. and Bulik, T. and Bulten, H. J. and Buskulic, D. and Cagnoli, G. and Calloni, E. and Canepa, M. and Carapella, G. and Carbognani, F. and Carullo, G. and Casanueva Diaz, J. and Casentini, C. and Casta\~neda, J. and Caudill, S. and Cavalier, F. and Cavalieri, R. and Cella, G. and Cerd\'a-Dur\'an, P. and Cesarini, E. and Chaibi, O. and Chassande-Mottin, E. and Chiadini, F. and Chierici, R. and Chincarini, A. and Chiummo, A. and Christensen, N. and Chua, S. and Ciani, G. and Ciecielag, P. and Cie\ifmmode \acute{s}\else \'{s}\fi{}lar, M. and Ciolfi, R. and Cipriano, F. and Cirone, A. and Clesse, S. and Cleva, F. and Coccia, E. and Cohadon, P.-F. and Cohen, D. and Colpi, M. and Conti, L. and Cordero-Carri\'on, I. and Corezzi, S. and Corre, D. and Cortese, S. and Coulon, J.-P. and Croquette, M. and Cudell, J.-R. and Cuoco, E. and Curylo, M. and D'Angelo, B. and D'Antonio, S. and Dattilo, V. and Davier, M. and Degallaix, J. and De Laurentis, M. and Del\'eglise, S. and Del Pozzo, W. and De Pietri, R. and De Rosa, R. and De Rossi, C. and Dietrich, T. and Di Fiore, L. and Di Giorgio, C. and Di Giovanni, F. and Di Giovanni, M. and Di Girolamo, T. and Di Lieto, A. and Di Pace, S. and Di Palma, I. and Di Renzo, F. and Drago, M. and Ducoin, J.-G. and Durante, O. and D'Urso, D. and Eisenmann, M. and Errico, L. and Estevez, D. and Fafone, V. and Farinon, S. and Feng, F. and Fenyvesi, E. and Ferrante, I. and Fidecaro, F. and Fiori, I. and Fiorucci, D. and Fittipaldi, R. and Fiumara, V. and Flaminio, R. and Font, J. A. and Fournier, J.-D. and Frasca, S. and Frasconi, F. and Frey, V. and Fronz\`e, G. and Garufi, F. and Gemme, G. and Genin, E. and Gennai, A. and Ghosh, Archisman and Giacomazzo, B. and Gosselin, M. and Gouaty, R. and Grado, A. and Granata, M. and Greco, G. and Grignani, G. and Grimaldi, A. and Grimm, S. J. and Gruning, P. and Guidi, G. M. and Guix\'e, G. and Guo, Y. and Gupta, P. and Halim, O. and Harder, T. and Harms, J. and Heidmann, A. and Heitmann, H. and Hello, P. and Hemming, G. and Hennes, E. and Hinderer, T. and Hofman, D. and Huet, D. and Hui, V. and Idzkowski, B. and Iess, A. and Intini, G. and Isac, J.-M. and Jacqmin, T. and Jaranowski, P. and Jonker, R. J. G. and Katsanevas, S. and K\'ef\'elian, F. and Khan, I. and Khetan, N. and Koekoek, G. and Koley, S. and Kr\'olak, A. and Kutynia, A. and Laghi, D. and Lamberts, A. and La Rosa, I. and Lartaux-Vollard, A. and Lazzaro, C. and Leaci, P. and Leroy, N. and Letendre, N. and Linde, F. and Llorens-Monteagudo, M. and Longo, A. and Lorenzini, M. and Loriette, V. and Losurdo, G. and Lumaca, D. and Macquet, A. and Majorana, E. and Maksimovic, I. and Man, N. and Mangano, V. and Mantovani, M. and Mapelli, M. and Marchesoni, F. and Marion, F. and Marquina, A. and Marsat, S. and Martelli, F. and Martinez, V. and Masserot, A. and Mastrogiovanni, S. and Mejuto Villa, E. and Mereni, L. and Merzougui, M. and Metzdorff, R. and Miani, A. and Michel, C. and Milano, L. and Miller, A. and Milotti, E. and Minazzoli, O. and Minenkov, Y. and Montani, M. and Morawski, F. and Mours, B. and Muciaccia, F. and Nagar, A. and Nardecchia, I. and Naticchioni, L. and Neilson, J. and Nelemans, G. and Nguyen, C. and Nichols, D. and Nissanke, S. and Nocera, F. and Oganesyan, G. and Olivetto, C. and Pagano, G. and Pagliaroli, G. and Palomba, C. and Pang, P. T. H. and Pannarale, F. and Paoletti, F. and Paoli, A. and Pascucci, D. and Pasqualetti, A. and Passaquieti, R. and Passuello, D. and Patricelli, B. and Perego, A. and Pegoraro, M. and P\'erigois, C. and Perreca, A. and Perri\`es, S. and Phukon, K. S. and Piccinni, O. J. and Pichot, M. and Piendibene, M. and Piergiovanni, F. and Pierro, V. and Pillant, G. and Pinard, L. and Pinto, I. M. and Piotrzkowski, K. and Plastino, W. and Poggiani, R. and Popolizio, P. and Porter, E. K. and Prevedelli, M. and Principe, M. and Prodi, G. A. and Punturo, M. and Puppo, P. and Raaijmakers, G. and Radulesco, N. and Rapagnani, P. and Razzano, M. and Regimbau, T. and Rei, L. and Rettegno, P. and Ricci, F. and Riemenschneider, G. and Robinet, F. and Rocchi, A. and Rolland, L. and Romanelli, M. and Romano, R. and Rosi\ifmmode \acute{n}\else \'{n}\fi{}ska, D. and Ruggi, P. and Salafia, O. S. and Salconi, L. and Samajdar, A. and Sanchis-Gual, N. and Santos, E. and Sassolas, B. and Sauter, O. and Sayah, S. and Sentenac, D. and Sequino, V. and Sharma, A. and Sieniawska, M. and Singh, N. and Singhal, A. and Sipala, V. and Sordini, V. and Sorrentino, F. and Spera, M. and Stachie, C. and Steer, D. A. and Stratta, G. and Sur, A. and Swinkels, B. L. and Tacca, M. and Tanasijczuk, A. J. and Tapia San Martin, E. N. and Tiwari, S. and Tonelli, M. and Torres-Forn\'e, A. and Tosta e Melo, I. and Travasso, F. and Tringali, M. C. and Trovato, A. and Tsang, K. W. and Turconi, M. and Valentini, M. and van Bakel, N. and van Beuzekom, M. and van den Brand, J. F. J. and Van Den Broeck, C. and van der Schaaf, L. and Vardaro, M. and Vas\'uth, M. and Vedovato, G. and Verkindt, D. and Vetrano, F. and Vicer\'e, A. and Vinet, J.-Y. and Vocca, H. and Walet, R. and Was, M. and Zadro\ifmmode \dot{z}\else \.{z}\fi{}ny, A. and Zelenova, T. and Zendri, J.-P. and Vahlbruch, Henning and Mehmet, Moritz and L\"uck, Harald and Danzmann, Karsten},
  collaboration = {Virgo Collaboration},
  journal = {Phys. Rev. Lett.},
  volume = {123},
  issue = {23},
  pages = {231108},
  numpages = {10},
  year = {2019},
  month = {Dec},
  publisher = {American Physical Society},
  doi = {10.1103/PhysRevLett.123.231108},
  url = {https://link.aps.org/doi/10.1103/PhysRevLett.123.231108}
}

@article{RevModPhys.92.015004,
  title = {Sensitivity optimization for NV-diamond magnetometry},
  author = {Barry, John F. and Schloss, Jennifer M. and Bauch, Erik and Turner, Matthew J. and Hart, Connor A. and Pham, Linh M. and Walsworth, Ronald L.},
  journal = {Rev. Mod. Phys.},
  volume = {92},
  issue = {1},
  pages = {015004},
  numpages = {68},
  year = {2020},
  month = {Mar},
  publisher = {American Physical Society},
  doi = {10.1103/RevModPhys.92.015004},
  url = {https://link.aps.org/doi/10.1103/RevModPhys.92.015004}
}

@article{Ku-2020,
doi	= {10.1038/s41586-020-2507-2},
title	= {Imaging viscous flow of the Dirac fluid in graphene},
author	= {Ku, Mark J. H. and Zhou, Tony X. and Li, Qing and Shin, Young J. and Shi, Jing K. and Burch, Claire and Anderson, Laurel E. and Pierce, Andrew T. and Xie, Yonglong and Hamo, Assaf and Vool, Uri and Zhang, Huiliang and Casola, Francesco and Taniguchi, Takashi and Watanabe, Kenji and Fogler, Michael M. and Kim, Philip and Yacoby, Amir and Walsworth, Ronald L.},
journal	= {Nature},
year	= {2020},
day	= {22},
volume	= {583},
issue	= {7817},
page	= {537--541}
}

@article{Glenn-2018,
doi	= {10.1038/nature25781},
title	= {High-resolution magnetic resonance spectroscopy using a solid-state spin sensor},
author	= {Glenn, David R. and Bucher, Dominik B. and Lee, Junghyun and Lukin, Mikhail D. and Park, Hongkun and Walsworth, Ronald L.},
journal	= {Nature}, 
year	= {2018},
volume	= {555},
issue	= {7696},
page	= {351--354}
}

@article{PhysRevApplied.14.014097,
  title = {Magnetic Field Fingerprinting of Integrated-Circuit Activity with a Quantum Diamond Microscope},
  author = {Turner, Matthew J. and Langellier, Nicholas and Bainbridge, Rachel and Walters, Dan and Meesala, Srujan and Babinec, Thomas M. and Kehayias, Pauli and Yacoby, Amir and Hu, Evelyn and Lon\ifmmode \check{c}\else \v{c}\fi{}ar, Marko and Walsworth, Ronald L. and Levine, Edlyn V.},
  journal = {Phys. Rev. Appl.},
  volume = {14},
  issue = {1},
  pages = {014097},
  numpages = {15},
  year = {2020},
  month = {Jul},
  publisher = {American Physical Society},
  doi = {10.1103/PhysRevApplied.14.014097},
  url = {https://link.aps.org/doi/10.1103/PhysRevApplied.14.014097}
}

@article{RevModPhys.82.2313,
  title = {Quantum information with Rydberg atoms},
  author = {Saffman, M. and Walker, T. G. and M\o{}lmer, K.},
  journal = {Rev. Mod. Phys.},
  volume = {82},
  issue = {3},
  pages = {2313--2363},
  numpages = {0},
  year = {2010},
  month = {Aug},
  publisher = {American Physical Society},
  doi = {10.1103/RevModPhys.82.2313},
  url = {https://link.aps.org/doi/10.1103/RevModPhys.82.2313}
}

@article{Sedlacek-2012,
doi	= {10.1038/nphys2423},
title	= {Microwave electrometry with Rydberg atoms in a vapour cell using bright atomic resonances},
author	= {Sedlacek, Jonathon A. and Schwettmann, Arne and K\"ubler, Harald and L\"ow, Robert and Pfau, Tilman and Shaffer, James P.},
journal	= {Nature Physics1}, 
year	= {2012},
month	= {sep},
day	= {16},
volume	= {8},
issue	= {11},
pages	= {819--824}
}

@article{PhysRevLett.111.063001,
  title = {Atom-Based Vector Microwave Electrometry Using Rubidium Rydberg Atoms in a Vapor Cell},
  author = {Sedlacek, J. A. and Schwettmann, A. and K\"ubler, H. and Shaffer, J. P.},
  journal = {Phys. Rev. Lett.},
  volume = {111},
  issue = {6},
  pages = {063001},
  numpages = {5},
  year = {2013},
  month = {Aug},
  publisher = {American Physical Society},
  doi = {10.1103/PhysRevLett.111.063001},
  url = {https://link.aps.org/doi/10.1103/PhysRevLett.111.063001}
}

@article{Fan-2014,
doi	= {10.1364/ol.39.003030},
title	= {Subwavelength microwave electric-field imaging using Rydberg atoms inside atomic vapor cells},
author	= {Fan, H. Q. and Kumar, S. and Daschner, R. and K\"ubler, H. and Shaffer, J. P.},
journal	= {Optics Letters},
year	= {2014},
month	= {may},
day	= {14},
volume	= {39},
issue	= {10},
pages	= {3030}
}

@article{10.1063/1.5028357,
    author = {Meyer, David H. and Cox, Kevin C. and Fatemi, Fredrik K. and Kunz, Paul D.},
    title = {Digital communication with Rydberg atoms and amplitude-modulated microwave fields},
    journal = {Applied Physics Letters},
    volume = {112},
    number = {21},
    pages = {211108},
    year = {2018},
    month = {05},
    issn = {0003-6951},
    doi = {10.1063/1.5028357},
    url = {https://doi.org/10.1063/1.5028357},
    eprint = {https://pubs.aip.org/aip/apl/article-pdf/doi/10.1063/1.5028357/14511014/211108\_1\_online.pdf},
}

@article{10.1063/1.5099036,
    author = {Holloway, Christopher L. and Simons, Matthew T. and Haddab, Abdulaziz H. and Williams, Carl J. and Holloway, Maxwell W.},
    title = {A “real-time” guitar recording using Rydberg atoms and electromagnetically induced transparency: Quantum physics meets music},
    journal = {AIP Advances},
    volume = {9},
    number = {6},
    pages = {065110},
    year = {2019},
    month = {06},
    issn = {2158-3226},
    doi = {10.1063/1.5099036},
    url = {https://doi.org/10.1063/1.5099036}
}

@article{PhysRevX.10.011027,
  title = {Full-Field Terahertz Imaging at Kilohertz Frame Rates Using Atomic Vapor},
  author = {Downes, Lucy A. and MacKellar, Andrew R. and Whiting, Daniel J. and Bourgenot, Cyril and Adams, Charles S. and Weatherill, Kevin J.},
  journal = {Phys. Rev. X},
  volume = {10},
  issue = {1},
  pages = {011027},
  numpages = {7},
  year = {2020},
  month = {Feb},
  publisher = {American Physical Society},
  doi = {10.1103/PhysRevX.10.011027},
  url = {https://link.aps.org/doi/10.1103/PhysRevX.10.011027}
}

@article{PhysRevApplied.21.044025,
  title = {Polarization-insensitive microwave electrometry using Rydberg atoms},
  author = {Cloutman, Matthew and Chilcott, Matthew and Elliott, Alexander and Otto, J. Susanne and Deb, Amita B. and Kj\ae{}rgaard, Niels},
  journal = {Phys. Rev. Appl.},
  volume = {21},
  issue = {4},
  pages = {044025},
  numpages = {6},
  year = {2024},
  month = {Apr},
  publisher = {American Physical Society},
  doi = {10.1103/PhysRevApplied.21.044025},
  url = {https://link.aps.org/doi/10.1103/PhysRevApplied.21.044025}
}

@article{PhysRevA.96.012117,
  title = {Quantum parameter estimation with optimal control},
  author = {Liu, Jing and Yuan, Haidong},
  journal = {Phys. Rev. A},
  volume = {96},
  issue = {1},
  pages = {012117},
  numpages = {14},
  year = {2017},
  month = {Jul},
  publisher = {American Physical Society},
  doi = {10.1103/PhysRevA.96.012117},
  url = {https://link.aps.org/doi/10.1103/PhysRevA.96.012117}
}

@article{PhysRevA.96.042114,
  title = {Control-enhanced multiparameter quantum estimation},
  author = {Liu, Jing and Yuan, Haidong},
  journal = {Phys. Rev. A},
  volume = {96},
  issue = {4},
  pages = {042114},
  numpages = {11},
  year = {2017},
  month = {Oct},
  publisher = {American Physical Society},
  doi = {10.1103/PhysRevA.96.042114},
  url = {https://link.aps.org/doi/10.1103/PhysRevA.96.042114}
}

@article{PhysRevA.101.012321,
  title = {Robust control of a not gate by composite pulses},
  author = {Dridi, G. and Mejatty, M. and Glaser, S. J. and Sugny, D.},
  journal = {Phys. Rev. A},
  volume = {101},
  issue = {1},
  pages = {012321},
  numpages = {8},
  year = {2020},
  month = {Jan},
  publisher = {American Physical Society},
  doi = {10.1103/PhysRevA.101.012321},
  url = {https://link.aps.org/doi/10.1103/PhysRevA.101.012321}
}

@article{PhysRevA.90.023411,
  title = {Optimal control of the signal-to-noise ratio per unit time for a spin-1/2 particle},
  author = {Lapert, M. and Ass\'emat, E. and Glaser, S. J. and Sugny, D.},
  journal = {Phys. Rev. A},
  volume = {90},
  issue = {2},
  pages = {023411},
  numpages = {6},
  year = {2014},
  month = {Aug},
  publisher = {American Physical Society},
  doi = {10.1103/PhysRevA.90.023411},
  url = {https://link.aps.org/doi/10.1103/PhysRevA.90.023411}
}

@article{KOBZAR2012142,
title = {Exploring the limits of broadband 90$^{\circ}$ and 180$^{\circ}$ universal rotation pulses},
journal = {Journal of Magnetic Resonance},
volume = {225},
pages = {142-160},
year = {2012},
issn = {1090-7807},
doi = {https://doi.org/10.1016/j.jmr.2012.09.013},
url = {https://www.sciencedirect.com/science/article/pii/S1090780712003126},
author = {Kyryl Kobzar and Sebastian Ehni and Thomas E. Skinner and Steffen J. Glaser and Burkhard Luy}
}

@article{PhysRevLett.96.010401,
  title = {Quantum Metrology},
  author = {Giovannetti, Vittorio and Lloyd, Seth and Maccone, Lorenzo},
  journal = {Phys. Rev. Lett.},
  volume = {96},
  issue = {1},
  pages = {010401},
  numpages = {4},
  year = {2006},
  month = {Jan},
  publisher = {American Physical Society},
  doi = {10.1103/PhysRevLett.96.010401},
  url = {https://link.aps.org/doi/10.1103/PhysRevLett.96.010401}
}

@article{AdvancesQuantumMetrology_2011,
author = { Giovannetti, Vittorio and Lloyd, Seth and Maccone, Lorenzo},
title = {Advances in quantum metrology},
journal = {Nature Photonics},
year = {2011},
volume  = {5},
pages = {222},
doi = { 10.1038/nphoton.2011.35}
}

@article{PhysRevA.70.013402,
  title = {Stabilization of ultracold molecules using optimal control theory},
  author = {Koch, Christiane P. and Palao, Jos\'e P. and Kosloff, Ronnie and Masnou-Seeuws, Fran\ifmmode \mbox{\c{c}}\else \c{c}\fi{}oise},
  journal = {Phys. Rev. A},
  volume = {70},
  issue = {1},
  pages = {013402},
  numpages = {14},
  year = {2004},
  month = {Jul},
  publisher = {American Physical Society},
  doi = {10.1103/PhysRevA.70.013402},
  url = {https://link.aps.org/doi/10.1103/PhysRevA.70.013402}
}

@article{PRXQuantum.2.030203,
  title = {Introduction to the Pontryagin Maximum Principle for Quantum Optimal Control},
  author = {Boscain, U. and Sigalotti, M. and Sugny, D.},
  journal = {PRX Quantum},
  volume = {2},
  issue = {3},
  pages = {030203},
  numpages = {31},
  year = {2021},
  month = {Sep},
  publisher = {American Physical Society},
  doi = {10.1103/PRXQuantum.2.030203},
  url = {https://link.aps.org/doi/10.1103/PRXQuantum.2.030203}
}

@article{PRXQuantum.2.010101,
  title = {From Pulses to Circuits and Back Again: A Quantum Optimal Control Perspective on Variational Quantum Algorithms},
  author = {Magann, Alicia B. and Arenz, Christian and Grace, Matthew D. and Ho, Tak-San and Kosut, Robert L. and McClean, Jarrod R. and Rabitz, Herschel A. and Sarovar, Mohan},
  journal = {PRX Quantum},
  volume = {2},
  issue = {1},
  pages = {010101},
  numpages = {16},
  year = {2021},
  month = {Jan},
  publisher = {American Physical Society},
  doi = {10.1103/PRXQuantum.2.010101},
  url = {https://link.aps.org/doi/10.1103/PRXQuantum.2.010101}
}

@article{doi:10.1116/5.0006785,
author = {Rembold, Phila  and Oshnik, Nimba  and M\"uller, Matthias M.  and Montangero, Simone  and Calarco, Tommaso  and Neu, Elke },
title = {Introduction to quantum optimal control for quantum sensing with nitrogen-vacancy centers in diamond},
journal = {AVS Quantum Science},
volume = {2},
number = {2},
pages = {024701},
year = {2020},
doi = {10.1116/5.0006785},
URL = { https://doi.org/10.1116/5.0006785}
}

@misc{ansel2024introduction,
      title={Introduction to Theoretical and Experimental aspects of Quantum Optimal Control}, 
      author={Q. Ansel and E. Dionis and F. Arrouas and B. Peaudecerf and S. Guérin and D. Gu\'ery-Odelin and D. Sugny},
      year={2024},
      eprint={arXiv: 2403.00532},
      archivePrefix={arXiv},
      primaryClass={quant-ph}
}

@article{PhysRevA.82.063422,
  title = {Frictionless atom cooling in harmonic traps: A time-optimal approach},
  author = {Stefanatos, Dionisis and Ruths, Justin and Li, Jr-Shin},
  journal = {Phys. Rev. A},
  volume = {82},
  issue = {6},
  pages = {063422},
  numpages = {8},
  year = {2010},
  month = {Dec},
  publisher = {American Physical Society},
  doi = {10.1103/PhysRevA.82.063422},
  url = {https://link.aps.org/doi/10.1103/PhysRevA.82.063422}
}

@article{PhysRevA.87.043607,
  title = {Cooling through optimal control of quantum evolution},
  author = {Rahmani, Armin and Kitagawa, Takuya and Demler, Eugene and Chamon, Claudio},
  journal = {Phys. Rev. A},
  volume = {87},
  issue = {4},
  pages = {043607},
  numpages = {6},
  year = {2013},
  month = {Apr},
  publisher = {American Physical Society},
  doi = {10.1103/PhysRevA.87.043607},
  url = {https://link.aps.org/doi/10.1103/PhysRevA.87.043607}
}

@article{doi:10.1137/100818431,
author = {Stefanatos, Dionisis and Schaettler, Heinz and Li, Jr-Shin},
title = {Minimum-Time Frictionless Atom Cooling in Harmonic Traps},
journal = {SIAM Journal on Control and Optimization},
volume = {49},
number = {6},
pages = {2440-2462},
year = {2011},
doi = {10.1137/100818431},
URL = { https://doi.org/10.1137/100818431    }
}

@article{PhysRev.78.695,
  title = {A Molecular Beam Resonance Method with Separated Oscillating Fields},
  author = {Ramsey, Norman F.},
  journal = {Phys. Rev.},
  volume = {78},
  issue = {6},
  pages = {695--699},
  numpages = {0},
  year = {1950},
  month = {Jun},
  publisher = {American Physical Society},
  doi = {10.1103/PhysRev.78.695},
  url = {https://link.aps.org/doi/10.1103/PhysRev.78.695}
}

@article{KHANEJA2005296,
title = "Optimal control of coupled spin dynamics: design of NMR pulse sequences by gradient ascent algorithms",
journal = "Journal of Magnetic Resonance",
volume = "172",
number = "2",
pages = "296 - 305",
year = "2005",
issn = "1090-7807",
doi = "https://doi.org/10.1016/j.jmr.2004.11.004",
url = "http://www.sciencedirect.com/science/article/pii/S1090780704003696",
author = "Navin Khaneja and Timo Reiss and Cindie Kehlet and Thomas Schulte-Herbrüggen and Steffen J. Glaser"
}

@article{PhysRevA.101.022320,
  title = {Time-optimal control of a dissipative qubit},
  author = {Lin, Chungwei and Sels, Dries and Wang, Yebin},
  journal = {Phys. Rev. A},
  volume = {101},
  issue = {2},
  pages = {022320},
  numpages = {12},
  year = {2020},
  month = {Feb},
  publisher = {American Physical Society},
  doi = {10.1103/PhysRevA.101.022320},
  url = {https://link.aps.org/doi/10.1103/PhysRevA.101.022320}
}

@article{PhysRevA.103.052607,
  title = {Optimal control for quantum metrology via Pontryagin's principle},
  author = {Lin, Chungwei and Ma, Yanting and Sels, Dries},
  journal = {Phys. Rev. A},
  volume = {103},
  issue = {5},
  pages = {052607},
  numpages = {8},
  year = {2021},
  month = {May},
  publisher = {American Physical Society},
  doi = {10.1103/PhysRevA.103.052607},
  url = {https://link.aps.org/doi/10.1103/PhysRevA.103.052607}
}

@article{PhysRevA.105.042621,
  title = {Application of Pontryagin's maximum principle to quantum metrology in dissipative systems},
  author = {Lin, Chungwei and Ma, Yanting and Sels, Dries},
  journal = {Phys. Rev. A},
  volume = {105},
  issue = {4},
  pages = {042621},
  numpages = {10},
  year = {2022},
  month = {Apr},
  publisher = {American Physical Society},
  doi = {10.1103/PhysRevA.105.042621},
  url = {https://link.aps.org/doi/10.1103/PhysRevA.105.042621}
}

@article{PhysRevA.111.042602,
  title = {Time-optimal single-scalar control on a qubit of unitary dynamics},
  author = {Lin, Chungwei and Ding, Qi and Boufounos, Petros T. and Ma, Yanting and Wang, Yebin and Sels, Dries and Chien, Chih-Chun},
  journal = {Phys. Rev. A},
  volume = {111},
  issue = {4},
  pages = {042602},
  numpages = {15},
  year = {2025},
  month = {Apr},
  publisher = {American Physical Society},
  doi = {10.1103/PhysRevA.111.042602},
  url = {https://link.aps.org/doi/10.1103/PhysRevA.111.042602}
}

@article{PhysRevA.69.062320,
  title = {Cavity quantum electrodynamics for superconducting electrical circuits: An architecture for quantum computation},
  author = {Blais, Alexandre and Huang, Ren-Shou and Wallraff, Andreas and Girvin, S. M. and Schoelkopf, R. J.},
  journal = {Phys. Rev. A},
  volume = {69},
  issue = {6},
  pages = {062320},
  numpages = {14},
  year = {2004},
  month = {Jun},
  publisher = {American Physical Society},
  doi = {10.1103/PhysRevA.69.062320},
  url = {https://link.aps.org/doi/10.1103/PhysRevA.69.062320}
}

@article{Lupascu-2007,
doi	= {10.1038/nphys509},
title	= {Quantum non-demolition measurement of a superconducting two-level system},
author	= {A. Lupascu and S. Saito and T. Picot and P.C. de Groot and C. J. P. M. Harmans and J. E. Mooij},
journal	= {Nature Physics},
year	= {2007},
month	= {jan},
day	= {14},
volume	= {3},
issue	= {2},
page	= {119--123}
}

@article{RevModPhys.93.025005,
  title = {Circuit quantum electrodynamics},
  author = {Blais, Alexandre and Grimsmo, Arne L. and Girvin, S. M. and Wallraff, Andreas},
  journal = {Rev. Mod. Phys.},
  volume = {93},
  issue = {2},
  pages = {025005},
  numpages = {72},
  year = {2021},
  month = {May},
  publisher = {American Physical Society},
  doi = {10.1103/RevModPhys.93.025005},
  url = {https://link.aps.org/doi/10.1103/RevModPhys.93.025005}
}


\end{document}